\documentclass[%
superscriptaddress,
preprint,
showpacs,preprintnumbers,
nofootinbib,
 amsmath,amssymb,
 aps,
prc
]{revtex4-1}
\usepackage{float}
\usepackage{subfig}
\usepackage{feynmf}
\usepackage{graphicx}
\usepackage{dcolumn}
\usepackage{bm}
\usepackage{slashed}
\usepackage[usenames]{xcolor}
\usepackage{appendix}
\usepackage{braket}
\newcolumntype{P}[1]{>{\centering\arraybackslash}p{#1}}

\newcommand{\nsat}{n_{\mathrm{sat}}}
\newcommand{\keV}{\,\mathrm{keV}}
\newcommand{\MeV}{\,\mathrm{MeV}}
\newcommand{\GeV}{\,\mathrm{GeV}}

\newcommand{\kF}{k_\text{F}}

\newcommand{\tr}{\mathrm{Tr}}
\newcommand{\beq}{\begin{equation}}
\newcommand{\eeq}{\end{equation}}
\newcommand{\beqa}{\begin{eqnarray}}
\newcommand{\eeqa}{\end{eqnarray}}

\newcommand{\mycomment}[1]{}

\begin{document}

\title{The Kohn-Luttinger effect in dense matter and its implications for neutron stars.}
\date{\today}
\author{Mia Kumamoto}
\email{mialk@uw.edu}
\affiliation{Institute for Nuclear Theory, University of Washington, Seattle, WA USA}
\affiliation{Department of Physics, University of Washington, Seattle, WA USA}
\author{Sanjay Reddy}
\email{sareddy@uw.edu}
\affiliation{Institute for Nuclear Theory, University of Washington, Seattle, WA USA}

\begin{abstract}
Repulsive short-range interactions can induce p-wave attraction between fermions in dense matter and lead to Cooper pairing at the Fermi surface. We investigate this phenomenon, well-known as the Kohn-Luttinger effect in condensed matter physics, in dense matter with strong short-range repulsive interactions. We find that repulsive interactions required to stabilize massive neutron stars can induce p-wave pairing in neutron and quark matter. When massive vector bosons mediate the interaction between fermions, the induced interaction favors Cooper pairing in the $^3P_2$ channel. For the typical strength of the interaction favored by massive neutron stars, the associated pairing gaps in neutrons can be in the range of 10 keV to 10 MeV. Strong and attractive spin-orbit and tensor forces between neutrons can result in repulsive induced interactions that greatly suppress the $^3P_2$ pairing gap in neutron matter.  In quark matter, the induced interaction is too small to result in pairing gaps of phenomenological relevance.   
\end{abstract} 
\maketitle

\section{Introduction}
\label{sec:Introduction}
The discovery of massive neutron stars by radio observations of pulsars \cite{NANOGrav:2019jur,Antoniadis:2013pzd,Demorest:2010bx} confirmed that the maximum mass of neutron stars $M_{\mathrm {max}}>2~M_\odot$, and gravitational wave and x-ray observations constrain the radius of a neutron star with mass $~\simeq 1.4~M_\odot$ to the range $11-13$ km \cite{LIGOScientific:2018cki,De:2018uhw,Capano:2019eae,Miller:2019cac,Riley:2019yda}. These constraints and theoretical calculations of the EOS of neutron-rich matter at $n_B \lesssim 2 \nsat$ \cite{Hebeler:2009iv,Gandolfi:2013baa,Tews:2012fj,Lynn:2015jua,Drischler:2020hwi}, taken together strongly suggest a rapid increase in the pressure and the speed of sound in the NS core \cite{Tews:2018kmu}. This, in turn, implies strong repulsive interactions are necessary for any putative phase of high-density matter in the core. This article addresses whether such repulsion can have other observable consequences. In particular, we investigate if such repulsion can lead to  Cooper pairing between fermions with non-zero angular momentum due to the Kohn-Luttinger(KL) effect \cite{KohnLuttinger1965PRL} in the cores of neutron stars. 

The KL effect, which arises because the interaction at the Fermi surface is modified due to screening in the medium, implies that the Cooper pairing instability in high angular momentum states is inevitable and occurs even when the bare interaction is repulsive \cite{KohnLuttinger1965PRL}. The effect has been discussed extensively in condensed matter physics (For a recent pedagogic review, see Ref.~\cite{Kagan:2013}). In the context of dense nuclear matter, early work in \cite{Fay:1968,Pines:1971,Clark:1976} recognized that the interaction between nucleons induced by polarization effects in the medium would significantly alter the pairing gaps (for recent reviews, see \cite{Dean:2002zx, Gezerlis:2014efa}). The induced interaction, typically calculated in second-order perturbation theory or Fermi liquid theory,  naturally incorporates the KL effect. In dilute Fermi systems with attractive s-wave short-range interactions, it has been known since the work of Gor’kov and Melik-Barkhudarov that the induced interaction suppresses the s-wave pairing gap relative to the BCS prediction \cite{Gorkov:1961}. In neutron matter, when the s-wave interaction is repulsive, the induced interaction was initially expected to increase the p-wave attraction between neutrons\cite{Fay:1968}. However, more recent work in Ref.~\cite{Schwenk:2003bc} finds that the induced spin-orbit interaction can dominate and result in a net suppression instead at modest density. Here, we revisit calculating the induced interaction in high-density matter characterized by a large sound speed to study its implications for $^3P_2$ pairing. We consider short-range interactions that contain central and non-central components and study the competition between the attractive and repulsive components of the induced p-wave interaction and its density dependence. 

In quark matter, when the Fermi surfaces of up, down, and strange quarks are split due to charge neutrality and a larger strange quark mass, the KL effect provides a mechanism to pair quarks of the same flavor and color. However, in this case, we find that p-wave interaction induced by short-range repulsion introduced to increase the pressure of quark matter is too small to be of phenomenological relevance. 

Our study, which relies on extrapolating results derived from perturbation theory to strong coupling, provides order-of-magnitude estimates for the pairing gaps. Although the method we employ is inadequate to make quantitative predictions, it identifies a mechanism for $^3P_2$ pairing in dense Fermi systems with large repulsive interactions mediated by short-range interactions mediated by heavy vector bosons.

In section \ref{sec:KL_mechanism}, we review the KL mechanism for non-relativistic fermions. In section \ref{sec:nmatter}, we derive the induced interaction between neutrons at high density by assuming that the bare interaction is due to the exchange of heavy vector mesons.  In section \ref{sec:qmatter}, we consider the possible effects of the KL mechanism in quark matter. We discuss the implications for neutron star cooling in section \ref{sec:implications}, summarize our main findings, and discuss open questions in section \ref{sec:Conclusions}.  

\section{Kohn-Luttinger Mechanism}
\label{sec:KL_mechanism}
Kohn and Luttinger showed that a short-range repulsive potential can induce attraction in large odd partial waves due to medium effects that can overscreen the effective interaction between fermions at finite density \cite{KohnLuttinger1965PRL}. There has been renewed interest in studying the KL effect in condensed matter systems because calculations suggest that the induced pairing gaps in $p$-waves and low-order partial waves could be large enough to be realized in experiments (see, for example, \cite{BaranovKL,Gonzalez2008PRB,Nandkishoreetal2014PRB,Gonzalez2019PRL}).  KL's original calculation included terms at second order in the potential; more recent analysis \cite{Efremov_2000} calculates the potential up to fourth order in a constant potential characterized by a large scattering length as well as including retardation effects where pairing occurs away from the Fermi surface, also contributing at fourth order.

\begin{figure}[ht]
\caption{Irreducible second order diagrams for Kohn-Luttinger mechanism}
\vspace{4ex}
\label{fig:kldiagrams}
\begin{minipage}[b]{.3\linewidth}
    \begin{fmffile}{diagb}
        \begin{fmfgraph*}(80,80)
            \fmfstraight
            \fmftop{t1,t2}
            \fmfbottom{b1,b2}
            \fmfleft{l1}
            \fmfright{r1}
            \fmf{fermion}{b1,v1}
            \fmf{fermion}{v1,t1}
            \fmf{fermion}{b2,v4}
            \fmf{fermion}{v4,t2}
            \fmffreeze
            \fmf{photon,tension=0.2}{v1,v2}
            \fmf{fermion,left,label=$\ell+q$,tension=0.1}{v2,v3}
            \fmf{fermion,left,label=$\ell$,tension=0.1}{v3,v2}
            \fmf{photon,tension=0.2}{v3,v4}
            \fmfv{label=$-k;1$}{b1}
            \fmfv{label=$k;2$}{b2}
            \fmfv{label=$-k';3$}{t1}
            \fmfv{label=$k';4$}{t2}
        \end{fmfgraph*}
    \end{fmffile}
\end{minipage}
\begin{minipage}[b]{.3\linewidth}
    \begin{fmffile}{diagc}
        \begin{fmfgraph*}(80,80)
            \fmfstraight
            \fmfleft{l1,l2,l3,l4,l5}
            \fmfright{r1,r2,r3,r4,r5}
            \fmf{fermion}{l1,l3}
            \fmf{fermion}{l3,l5}
            \fmf{fermion}{r1,r2}
            \fmf{fermion}{r4,r5}
            \fmf{fermion,label=$\ell$,label.side=left}{r2,v1}
            \fmf{fermion,label=$\ell+q$,label.side=left}{v1,r4}
            \fmf{photon}{r2,r4}
            \fmf{photon}{l3,v1}
            \fmfv{label=$-k;1$}{l1}
            \fmfv{label=$-k';3$}{l5}
            \fmfv{label=$k;2$}{r1}
            \fmfv{label=$k';4$}{r5}
        \end{fmfgraph*}
    \end{fmffile}
\end{minipage}
\begin{minipage}[b]{.3\linewidth}
    \begin{fmffile}{diagd}
        \begin{fmfgraph*}(80,80)
            \fmfstraight
            \fmfleft{l1,l2,l5,l3,l4}
            \fmfright{r1,r2,r5,r3,r4}
            \fmf{fermion}{l1,l2}
            \fmf{fermion,label=$\ell+q$,label.side=left}{l2,l3}
            \fmf{fermion}{l3,l4}
            \fmf{fermion}{r1,r2}
            \fmf{fermion,label=$\ell$}{r2,r3}
            \fmf{fermion}{r3,r4}
            \fmf{photon}{r2,l3}
            \fmf{photon}{l2,r3}
            \fmfv{label=$-k;1$}{l1}
            \fmfv{label=$k';4$}{l4}
            \fmfv{label=$k;2$}{r1}
            \fmfv{label=$-k';3$}{r4}
        \end{fmfgraph*}
    \end{fmffile}
\end{minipage}
\vspace{4ex}
\end{figure}
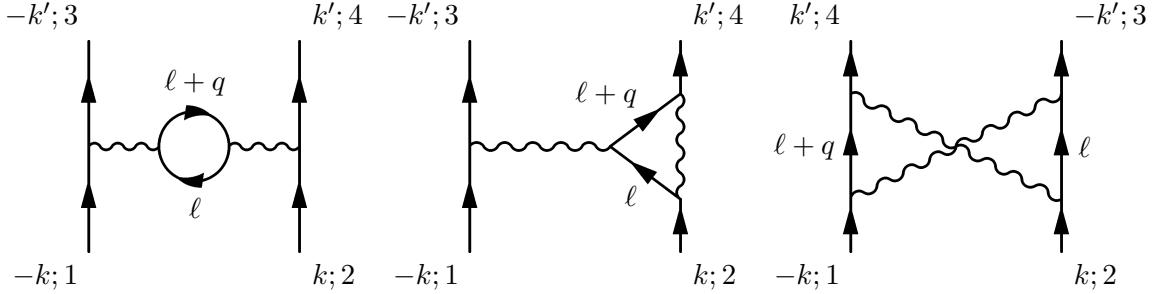

In weak coupling, the KL effect arises naturally at second order in the potential by evaluating the diagrams in Fig. \ref{fig:kldiagrams}.  We refer to these diagrams from left to right as the screening, vertex, and exchange diagrams, respectively. The vertex diagram also has a mirror image, which must be included. We consider interactions that occur at the Fermi surface, so $|k|=|k'|=k_F$.  The momentum transfer is labeled $q = k' - k$.

For a short-range potential with zero range and strength denoted by $U_0$, the non-relativistic calculation of these diagrams is straightforward. In this case, the screening diagram cancels the contribution from the two vertex diagrams, and only the exchange diagram contributes. The exchange diagram also gets an overall sign since it is crossed and is given by
\begin{equation}
    V_{KL}(q) = U_0^2 \frac{1}{\beta} \sum_{\ell_0} \int \frac{d^3 \ell}{(2\pi)^3} \frac{1}{\ell_0 - \ell^2/2m} \frac{1}{\ell_0 - (\ell+q)^2/2m}
\label{eq:vkl1}
\end{equation}
Notice that since $|k|=|k'|$, the frequency transfer $q_0$ is just zero.  Taking the Matsubara sum and simplifying it gives a singular contribution to the potential.  Since we are considering the effects of interactions with the medium, the loop integral has a factor $n_F(\ell^2/2m) = 1/(e^{\beta(\ell^2/2m-\mu)} + 1)$ and does not need to be regulated.  At the low temperatures we consider ($T \ll \epsilon_F=k_F^2/2m$) this simplifies $n_F(\ell^2/2m) \approx \Theta (k_F - \ell)$. The momentum integral in Eq.~\ref{eq:vkl1} yields the Lindhard function defined by 
\begin{equation}
\begin{split}
     U(q)= 
  -\frac{m}{4\pi^3 q} \int \ell \, d\ell \, d\Omega_\ell \frac{\Theta (k_F - \ell)}{\cos \theta_{q\ell} - q/2\ell}
  = \frac{mk_F}{4\pi^2} \left[ 1 - \frac{1}{\overline{q}} \left( 1- \frac{\overline{q}^2}{4} \right) \mbox{log} \left| \frac{1 - \overline{q}/2}{1+\overline{q}/2} \right| \right]
\end{split}
\end{equation}
where $\bar{q}=q/\kF$. Thus, Eq.~\ref{eq:vkl1} can be written as 
\begin{equation}
    \label{eq:nonrelpot}
V_{KL} (q) = -U_0^2 ~U(q)
\end{equation}

For any potential, the contribution to the induced potential from the singularity at $q = 2k_F$ of the Lindhard function scales as $(-1)^L L^{-4}$ for large L \cite{KohnLuttinger1965PRL, MaitiChubukov2013AIPKLReview} where $L$ is the angular momentum quantum number.  Since the regular  contributions to the total potential falls off exponentially with $L$, attraction is guaranteed 
for large odd partial waves.   

Although these results are only generically true for large $L$, they persist for relatively low partial waves for some potentials.  It was shown by \cite{BaranovKL, Efremov_2000} that the constant potential calculated above would result in p-wave attraction. The p-wave contribution from the potential in Eq.~\ref{eq:nonrelpot} can be found easily by making a change of integration variable from $\int_{-1}^1 d\cos \theta$ to $\int_0^2 \overline{q} d\overline{q}$. The matrix element in the Born approximation should be doubled due to a diagram with outgoing momenta switched, but we will absorb this normalization into the gap equation to match the literature.  The p-wave potential from the exchange diagram and its crossed counterpart is then given by
\begin{equation}
\label{eq:vl1NR}
    V_{\ell = 1} = -U_0^2 \frac{mk_F}{4\pi} \frac{4}{5\pi}(2 \log 2-1)
\end{equation}

The superfluid gap due to the induced attraction in p-waves was calculated several decades earlier in Ref.~\cite{Fay:1968,Kagan:1988}. In the BCS approximation, the p-wave gap 
\begin{equation}
\label{eq:DPNR}
\Delta_p \simeq  \epsilon_F ~\exp{\left( \frac{2}{N(0) V_{\ell = 1} } \right)} = \epsilon_F ~\exp{\left( \frac{-5 \pi^2}{4(2\ln{2}-1) (a\kF)^2} \right)}\,,     
\end{equation}
where $\epsilon_F=\kF^2/2m$ is the Fermi energy, $N(0)= m \kF / 2 \pi^2$ is the density of states at the Fermi surface for each spin, and the scattering length $a=m U_0/4\pi$ in weak coupling. 

In strong coupling, we cannot calculate the effective interaction at the Fermi surface reliably, and in the following, we shall assume that Eq.~\ref{eq:DPNR} provides a useful estimate. Further, we shall also assume that the s-wave scattering amplitude between quasi-particles at the Fermi surface, denoted by $f_0$, is directly related to the strength of the bare interaction $U_0$. In Fermi liquid theory (FLT), the sound speed 
\begin{equation}
\label{eq:cs2}
c_s = \frac{k_F}{\sqrt{3 m m^*}} \sqrt{(1+F_0)} \,, 
\end{equation}
where $F_0 = N(0) f_0$ is dimensionless measure of the quasi-particle interaction and $m^*$ is the fermion effective mass at the Fermi surface. Using this relation, we can estimate the interaction strength $U_0 \approx f_0 $ at a given density if $c_s$ and $m^*$ are known. 
If $ m^* \approx m $ and $U_0=f_0$, then the induced p-wave gap in Eq.~\ref{eq:DPNR} can be rewritten as 
\begin{equation}
\label{eq:DPNRF0}
\Delta_p \approx \epsilon_F ~\exp{\left( -\frac{5}{(2 \ln{(2)}-1) F_0^2} \right)}\,,     
\end{equation}
to illustrate its extreme sensitivity to $F_0$ and the sound speed through Eq.~\ref{eq:cs2}. For example, models of high-density neutron matter typically predict $F_0 \gtrsim 2 $ for $n_B \gtrsim 3~\nsat$ \cite{Friman:2019ncm}. Under these conditions, Eq.\ref{eq:DPNRF0} predicts robust p-wave pairing with gaps $\Delta_p \gtrsim 1$ MeV due to the induced interaction.  

In the next section, we will calculate the induced interaction between neutrons in more realistic scenarios where the bare potential is momentum-dependent and contains central and non-central components. 

\section{Induced p-wave pairing in dense neutron matter}
\label{sec:nmatter}

The s-wave potential at the Fermi surface becomes repulsive in the neutron star core when $n_B\gtrsim \nsat/2$. At these higher densities, $^3P_2$ pairing is favored because the bare potential in this channel remains attractive, and non-central components of the interaction, especially the spin-orbit interaction, favor the alignment of spin and orbital angular momentum. Calculations of the $^3P_2$ pairing gap in the BCS approximation reported in Refs.\cite{Baldoetal1998PRCNeutronGaps,Gezerlis:2014efa} show that the pairing gaps are model dependent, especially for $n_B > 2 \nsat$ because the nucleon-nucleon potentials at the relevant momenta are not well constrained by scattering data. In these calculations, the maximum value of the gap $\Delta_{^3P_2} \simeq 1-2$ MeV occurs between $2-3~\nsat$ and decreases rapidly with increasing density. At lower density, when the nucleon momenta 
$p \ll \Lambda_\chi$ where $\Lambda_\chi \simeq 500$ MeV is the breakdown scale of chiral EFT, a recent study used chiral EFT potentials and found that the maximum value $\Delta_{^3P_2} \simeq 0.4$ MeV was reached at $n_B \simeq 1.3 ~\nsat$ and its decrease at higher density was found to be sensitive to the details of the short-distance physics \cite{Drischler:2016cpy}. Together, these findings suggest that if neutron matter persists at the highest densities encountered in neutron stars, the bare $^3P_2$ potential could be small, and $\Delta_{^3P_2}$ depends on model assumptions about the nuclear interaction at short distances.   

When the bare $^3P_2$ potential weakens, the gap is especially sensitive to corrections due to induced interactions in the medium (see, for example, the discussion in section 3.4 of Ref.~\cite{Gezerlis:2014efa}). In early work, the interaction induced by the central components of the nuclear force was found to increase the $^3P_2$ gap \cite{Fay:1968,Pethick:1991}, as would be expected from the discussion of the KL mechanism in section \ref{sec:KL_mechanism}. However, as mentioned earlier, calculations that employed realistic low-energy nuclear potentials with significant non-central components found that the interference between the central and spin-orbit component of the nuclear force led to significant suppression of the $^3P_2$ gap for $n_B< 2 \nsat$ \cite{Schwenk:2003bc}. 

At $n_B \gtrsim 2 \nsat$, the description of nuclear interactions relies on model assumptions since the typical nucleon momenta $p \gtrsim \Lambda_\chi$.  To investigate the competition between a strong and repulsive central force and the spin-orbit component of the nuclear force at high density, we revisit the calculation of the induced interaction in simple models.  In what follows, we shall assume that the dominant contribution to the nucleon-nucleon interaction at short distances is due to the exchange of heavy vector mesons such as the $\omega$ and $\rho$ mesons with masses $m_\omega \approx m_\rho \simeq 800$ MeV. For $n_B \lesssim 4\nsat$, $\kF/\Lambda$ where $\Lambda \simeq m_\omega$ remains a useful expansion parameter. In this case, including terms up to ${\cal O}[(\kF/\Lambda)^2]$, the interaction can be described by the potential  
\begin{equation}
\label{eq:vnnso}
\begin{split}
V(q,q')&= C_0 + \tilde{C}_0 \sigma_1\cdot \sigma_2 + C_2 (q^2 + q'^2) + C'_2 (q'^2-q^2)  \\ 
&+  \left[\tilde{C}_2 (q^2 + q'^2)  + \tilde{C}'_2 (q'^2-q^2)\right] \sigma_1\cdot \sigma_2 + i V_{SO}~ q \times q' \cdot (\sigma_1 + \sigma_2) \\
&+ V_T q \cdot \sigma_1 q \cdot \sigma_2\,.
\end{split}
\end{equation}
In neutron matter, due to the Pauli principle, only the combinations $\bar{C}_0=C_0-3\tilde{C}_0$, $\bar{C}_2=C_2-3\tilde{C}_2$, and $\bar{C}'_2=C'_2+\tilde{C}'_2$  are relevant.  In the full expansion of the vector meson potential, there is also a term proportional to $(q \times q' \cdot \sigma_1)(q \times q' \cdot \sigma_2)$ and higher powers of momentum in the central and spin-orbit interactions. An exchange tensor term proportional to $q' \cdot \sigma_1 q' \cdot \sigma_2$ is also allowed by the symmetries of the interaction but is not present in the vector exchange. In this exploratory study, we shall neglect the spin-orbit squared and tensor exchange components and truncate the potential at order $k_F^2$. In this case, 5 LECs denoted by $\bar{C}_0$, $\bar{C}_2$, $\bar{C}'_2$, $V_{SO}$, and $V_T$ are adequate. The large and attractive spin-orbit interaction, whose strength is set by $V_{SO}$ plays an important role, as discussed below. Since we consider incoming and outgoing momenta at the Fermi surface with zero center-of-mass momentum, $q=k_1-k_3$ and $q'=k_1-k_4$ with the momenta of neutrons in the initial state are given by $k_1$ and $k_2$, and the final state momenta are $k_3$ and $k_4$. 

\begin{figure}[ht]
\vskip 10pt
\begin{centering}
\begin{fmffile}{first-diagram}
\begin{fmfgraph*}(80,80)
            \fmfstraight
            \fmftop{t1,t2}
            \fmfbottom{b1,b2}
            \fmfleft{l1}
            \fmfright{r1}
            \fmf{fermion}{b1,v1}
            \fmf{fermion}{v1,t1}
            \fmf{fermion}{b2,v2}
            \fmf{fermion}{v2,t2}
            \fmf{fermion,left,label=$\ell+q;b$,tension=0.4}{v1,v2}
            \fmf{fermion,left,label=$\ell;a$,tension=0.4}{v2,v1}
            \fmfv{decoration.shape=circle,decoration.filled=shaded,label=$V_L$,label.dist=40}{v1}
            \fmfv{decoration.shape=circle,decoration.filled=shaded,label=$V_R$,label.dist=40}{v2}
            \fmfv{label=$-k;1$}{b1}
            \fmfv{label=$k;2$}{b2}
            \fmfv{label=$-k';3$}{t1}
            \fmfv{label=$k';4$}{t2}
        \end{fmfgraph*}
    \end{fmffile}\hspace{1in}
\begin{fmffile}{second-diagram}
\begin{fmfgraph*}(80,80)
            \fmfstraight
            \fmftop{t1,t2}
            \fmfbottom{b1,b2}
            \fmfleft{l1}
            \fmfright{r1}
            \fmf{fermion}{b1,v1}
            \fmf{fermion}{v1,t1}
            \fmf{fermion}{b2,v2}
            \fmf{fermion}{v2,t2}
            \fmf{fermion,left,label=$\ell+q';b$,tension=0.4}{v1,v2}
            \fmf{fermion,left,label=$\ell;a$,tension=0.4}{v2,v1}
            \fmfv{decoration.shape=circle,decoration.filled=shaded,label=$\bar{V}_L$,label.dist=40}{v1}
            \fmfv{decoration.shape=circle,decoration.filled=shaded,label=$\bar{V}_R$,label.dist=40}{v2}
            \fmfv{label=$-k;1$}{b1}
            \fmfv{label=$k;2$}{b2}
            \fmfv{label=$k';4$}{t1}
            \fmfv{label=$-k';3$}{t2}
        \end{fmfgraph*}
    \end{fmffile}  
\caption{ZS (left) and ZS' (right) diagrams. The hatched blobs represent the anti-symmetrized interaction defined in Eq.~\ref{eq:antisymm}.}
\label{fig:antisymmdiagram}
\end{centering}
\end{figure}
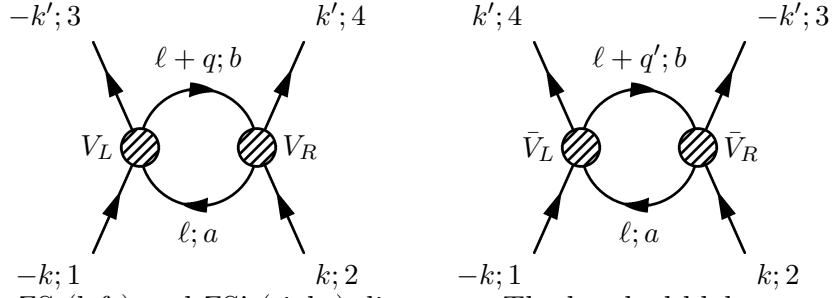

It is straightforward to repeat the calculation of the diagrams described in the preceding section with the potential in Eq.~\ref{eq:vnnso}. However, it is simpler to define the anti-symmetrized potential 
\begin{equation}
    \begin{split}
        V(q,q') &= C_0 (\delta_{13} \delta_{24} - \delta_{14} \delta_{23}) + C_2 (q^2 + q'^2) (\delta_{13} \delta_{24} - \delta_{14} \delta_{23}) \\
        &+ C'_2 (q'^2-q^2) (\delta_{13} \delta_{24} + \delta_{14} \delta_{23}) + \tilde{C}_0 (\sigma_{13} \cdot \sigma_{24} - \sigma_{14} \cdot \sigma_{23}) \\
        &+ \tilde{C}_2 (q^2 + q'^2) (\sigma_{13} \cdot \sigma_{24} - \sigma_{14} \cdot \sigma_{23}) + \tilde{C}'_2 (q'^2-q^2) (\sigma_{13} \cdot \sigma_{24} + \sigma_{14} \cdot \sigma_{23}) \\
        &+ 2i V_{SO} q \times q' \cdot (\sigma_{13} \delta_{24} + \sigma_{24} \delta_{13}) + V_T (q \cdot \sigma_{13} q \cdot \sigma_{24} - q' \cdot \sigma_{14} q' \cdot \sigma_{23})
    \end{split}
    \label{eq:antisymm}
\end{equation}
that includes the effect of the exchange processes. We use the notation $\delta_{ij} = \chi_j^\dagger \chi_i$ and $\sigma_{ij} = \chi_j^\dagger \sigma \chi_i$ for incoming and outgoing two-component spinors $\chi_i$ and $\chi_j^\dagger$.  In this case,  the induced interaction at second-order is calculated by evaluating the diagrams depicted in Fig.~\ref{fig:antisymmdiagram}. 
The diagram on the left is called the zero-sound diagram and denoted as ZS, and the diagram on the right is called the exchange zero-sound diagram and is denoted by the symbol ZS'. A detailed derivation of the total induced interaction  $V_{ind}=V_{ind}^{ZS}-V_{ind}^{ZS'}$ is presented in Appendix~\ref{App:InducedInteraction}. 

First, we present the result obtained by neglecting the momentum-dependent components of the bare central interaction. In this case, the induced potential 
\begin{equation}
\begin{split}
    V^{\rm ind} &= -(C_0^2 + 3\tilde{C}_0^2)(U(q) \delta_{14} \delta_{23} - U(q') \delta_{13} \delta_{24}) 
    + 6C_0 \tilde{C}_0 (U(q) \delta_{13} \delta_{24} - U(q')\delta_{14} \delta_{23}) \\
    &+ (-\tilde{C}_0^2 + 2C_0 \tilde{C}_0)(\sigma_{13} \cdot \sigma_{24} - \sigma_{14} \cdot \sigma_{23})(U(q) + U(q')) \\
    &- 3 \tilde{C}_0^2 (\sigma_{13} \cdot \sigma_{24} + \sigma_{14} \cdot \sigma_{23}) (U(q) - U(q'))
\end{split}
\end{equation}
The s- and p-wave potentials given by this interaction are 
\begin{equation}
    \begin{split}
        V^{\rm ind}_{^1S_0}(0) = \bar{C}^2_0 \frac{mk_F}{3\pi^2}(2 \log{2} + 1) \\
        V^{\rm ind}_{^3P_J}(0) = -\bar{C}^2_0  \frac{mk_F}{5\pi^2} ( 2 \log{2}-1)\,. 
    \end{split}
\end{equation}
where again $\bar{C}_0= C_0 - 3 \tilde{C}_0$ is the momentum-independent bare $^1S_0$ potential. 

The calculation of the induced potential, including the momentum-dependent central interactions, is tedious, and the analytic results contain a large number of terms. Details of the intermediate expressions can be found in Appendix \ref{App:InducedInteraction}.

We find analytic results for the second-order induced potentials in the $s$ and $p$ induced potentials. The induced $^1S_0$ and $^3P_J$ potentials calculated to $\mathcal{O}(mk_F^3)$ are given by 
\begin{equation}
    \begin{split}
        V^{\rm ind}_{^1S_0}&= mk_F \bar{C}_0^2 \frac{1}{3\pi^2} (1 + 2\log{2}) + mk_F^3 [\bar{C}_0 \bar{C}_2 \frac{2}{3\pi^2}(5+4\log{2}) \\
        &+ \bar{C}_0 \bar{C}'_2 \frac{2}{5\pi^2}(7-4\log{2}) - \bar{C}_0 V_T \frac{16}{15\pi^2}(2 + \log 2)]\,,
    \end{split}
    \label{eq:master1S0}
\end{equation}
and 
\begin{equation}
    \begin{split}
        V^{\rm ind}_{^3P_J} &= mk_F \bar{C}_0^2 \frac{1}{5\pi^2} (1 - 2\log{2}) + mk_F^3 [\bar{C}_0 \bar{C}_2 \frac{2}{105\pi^2}(59-68\log{2}) \\
        &- \bar{C}_0 \bar{C}'_2 \frac{2}{105\pi^2}(29+52\log{2}) \\
        &+\bar{C}_0 V_T \frac{1}{105\pi^2} (91J^2-221J+50 +(224J^2-544J+220)\log 2)] \,,
    \end{split}
    \label{eq:master3PJ}
\end{equation}
respectively. 

When momentum dependence of the central interaction is neglected, the bare spin-orbit force does not contribute to the induced interaction, and the leading order dependence of the induced potential on $J$ is determined by the tensor interaction. See Appendix \ref{App:InducedInteraction} for a detailed discussion. 

These contributions have the same behavior as the leading order KL result. The contribution to the induced interaction in a particular partial wave arising from terms in the bare interaction that do not contribute to that partial wave is strongly influenced by the KL singularity at $q=2\kF$. For this reason, their contribution is suppressed relative to other terms in the interaction at the same order in the expansion. Notice, for example, that in the p-wave induced potential, the term proportional to $\bar{C}_0 \bar{C}'_2$ has a numerical factor five times larger than $\bar{C}_0 \bar{C}_2$ and fifteen times larger than $\bar{C}_0 V_T$ for $^3P_2$. This implies that the singularity at $q=2\kF$ does not play an essential role when $\bar{C}'_2$ is of modest size. By comparing the relevant terms in Eq.~\ref{eq:master3PJ}, we find that the KL singularity plays an essential role only when  $\bar{C}'_2  \ll  \bar{C}_0/(16 \kF^2)$. In what follows, we shall continue to use the term Kohn-Luttinger effect to refer to the induced interaction, but it should be borne in mind that the singularity at $q=2\kF$ does not play a dominant role for typical values of $\bar{C}'_2$ we explore in this study. 

Since the spin-orbit force is strong and important in the $^3P_2$ channel, we expect that it may contribute to the induced interaction even though it enters at a higher order in the momentum expansion. To investigate the impact of the spin-orbit coupling, we calculate a subset of the terms that contain the central and spin-orbit interactions at $\mathcal{O}(mk_F^5)$. These corrections to the induced potentials are given by:
\begin{equation}
\begin{split}
    V_{^1S_0}^{(5)} &= mk_F^5[\bar{C}_2^2 \frac{8}{315\pi^2}(277+96\log{2}) - \bar{C}_2^{\prime 2} \frac{8}{105\pi^2} (43 + 24\log{2}) \\
    &+ \bar{C}_2 \bar{C}'_2 \frac{32}{105\pi^2}(37 + 6 \log{2}) +V_{SO}^2 \frac{8}{35\pi^2}(17+16\log 2)] \, ,
\end{split}
\end{equation}
and 
\begin{equation}
\label{eq:order_kf5}
    \begin{split}
    V_{^3P_J}^{(5)} &= mk_F^5 [\bar{C}_2^2 \frac{16}{567\pi^2}(83-24\log{2}) +\bar{C}_2^{\prime 2} \frac{64}{567\pi^2} (34 - 3\log{2}) \\
    &- \bar{C}_2 \bar{C}'_2 \frac{16}{2835\pi^2}(523 + 204 \log{2}) +\bar{C}'_2 V_{SO} [J(J+1)-4]\frac{32}{945\pi^2} (43 + 24\log{2}) \\
    &+ V_T V_{SO} [J(J+1) - 4] \frac{16}{945\pi^2}(43+24\log 2) \\
    &+ V_{SO}^2 (7J^2-17J+10)\frac{32}{4725\pi^2}(43+24\log 2)]\, .
    \end{split}
\end{equation}
These are not all of the terms that contribute at $\mathcal{O}(mk_F^5)$.  Not included are terms proportional to $\bar{C}_2 V_T$, $\bar{C}'_2 V_T$, and $V_T^2$. 

From Eq.~\ref{eq:order_kf5}, we deduce that when the bare spin-orbit interaction is attractive, it could  enhance $^3P_2$ pairing if 
\begin{equation}
2\bar{C}'_2+V_T+\frac{4}{5} V_{SO} >0\,.
\end{equation}
The relation of this result to the suppression of $^3P_2$ pairing at low density due to spin-orbit interactions found in Ref. \cite{Schwenk:2003bc}, which employed a realistic low-momentum nucleon-nucleon potential fit to scattering data warrants further study. 

We consider three scenarios to study the implications of these results for dense neutron matter. Each scenario is defined by a specification of the LECs that appear in the bare potential defined in Eq.~\ref{eq:vnnso}. In scenario A, we shall assume that the exchange of heavy vector mesons mediates the interactions between neutrons. When the mass of the vector meson is large compared to the neutron Fermi momentum, and interaction is described by a current-current four-fermion Lagrangian 
\begin{equation}
\label{eq:RFL}
    \mathcal{L}_{\rm int} =- \frac{G_V}{2} (\overline{n} \gamma_\mu n) (\overline{n} \gamma^\mu n)  \,.
\end{equation}
Retaining only the leading terms in the $k/m_n$ expansion, the LECs appearing in Eqs.~\ref{eq:master1S0} and \ref{eq:master3PJ} are given by 
\begin{equation} 
\bar{C}_0=G_V, \quad \bar{C}_2=\frac{5G_V}{8m_n^2}, \quad \bar{C}'_2=\frac{3G_V}{8m_n^2}, \quad V_{SO} = -\frac{3G_V}{8m_n^2}, \quad V_{T}=\frac{G_V}{4m_n^2}
\end{equation}
Although the simple vector interaction cannot capture the complex nature of interactions between neutrons, which could involve a richer operator structure due to pion exchange and many nucleon forces, it is able to describe the qualitative aspects of the nucleon-nucleon interaction at high momentum; it predicts negative phase shifts in the $^1S_0$, $^3P_0$, and $^3P_1$ channels. The phase shift in the $^3P_2$ channel vanishes because $ V_{SO} = -\bar{C}'_2$, and the spin-orbit interaction exactly cancels the contribution from the central force. This aspect of short-range vector interactions that leads to a vanishing bare potential in the $^3P_2$ channel is a generic feature of any four-fermion interaction without derivative couplings since initial and final states constructed only from spin and helicity operators cannot be combined to form a tensor of rank greater than 1.  As a result, such an interaction cannot generate potentials in channels with $J\geq 2$ at the tree level. Including momentum dependence in the meson propagator, explicit derivative couplings (i.e., momentum dependence beyond that found in the Dirac spinors), or momentum dependence from loops lifts this restriction. 
\begin{figure}[h]
\begin{center}
\includegraphics[width=0.7\textwidth]{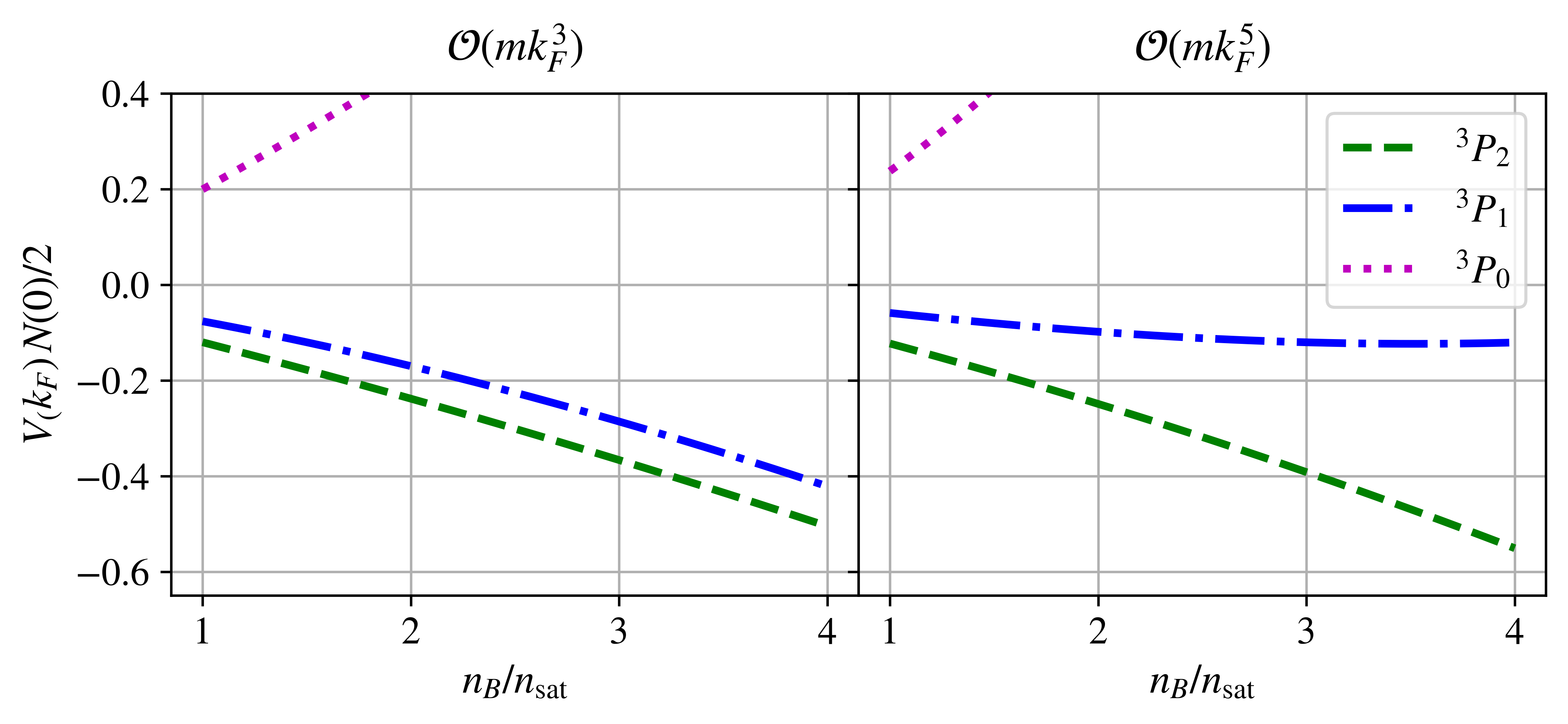}
\caption{Total potential for heavy vector boson exchange at $\mathcal{O}(mk_F^3)$ (left) and at $\mathcal{O}(mk_F^5)$ (right). The coupling constant is tuned to $F_0=3$ at $3\nsat$.}
\label{fig:vector_potentials}
\end{center}
\end{figure}

The couplings are related to the Fermi liquid parameters $F_0$ and $G_0$. In the mean-field theory, $F_0$ and $G_0$ depend only on the central components  of the interaction and are given by 
\begin{align}
\label{eq:F0G0}
F_0=N(0)\left(\bar{C}_0 + 2 \kF^2 (\bar{C}_2 + 3 \bar{C_2}')    \right)\,, \\
G_0=-N(0)\left(\bar{C}_0 + 2 \kF^2 (\bar{C}_2 - \bar{C_2}')    \right)\,, 
\end{align}
where $N(0)= \sqrt{\kF^2+m_n^2} \kF/2\pi^2$ is the density of states for each spin at the Fermi surface.
Thus, if $F_0$ and $G_0$ are specified, the strength of the s-wave components of the interaction are constrained by the equation  $(2\bar{C}_0 + 4 k_F^2 \bar{C}_2)=(F_0 - 3G_0)/2 N(0)$ and the p-wave component is obtained using the relation  $\bar{C}'_2 = (F_0 + G_0)/(8 N(0) k_F^2)$. 
In scenario A, the interaction contains just one parameter, $G_V$. In this case, $F_0$ and $G_0$ are not independent and $G_V$ is determined by specifying $F_0$, which we take to be in the range $2-4$ at $n_B= 3~\nsat$.

\begin{figure}[h]
\begin{center}
\includegraphics[width=0.7\textwidth]{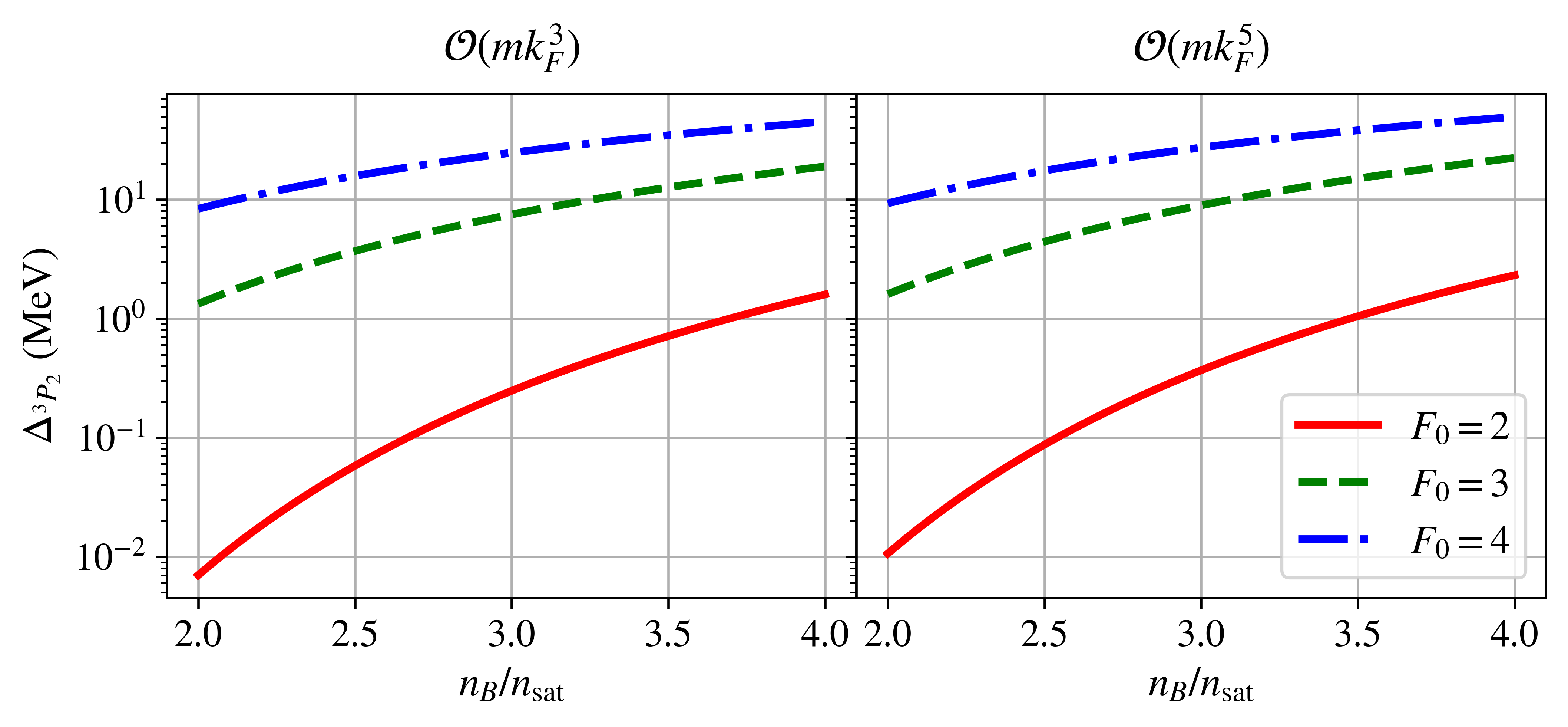}
\caption{The $^3P_2$ gap corresponding to the short-range vector interaction in Eq.~\ref{eq:RFL}. Results at $\mathcal{O}(mk_F^3)$ are shown in the left panel and at $\mathcal{O}(mk_F^5)$ in the right panel. The coupling constant is tuned to the given value of $F_0$ at $3\nsat$.}
\label{fig:vector_gaps}
\end{center}
\end{figure}
The induced and total p-wave interactions at the Fermi surface are shown in Fig.~\ref{fig:vector_potentials} for $F_0=3$ at $n_B= 3 \nsat$. The actual value shown $VN(0)/2$ is the quantity in the exponent of the BCS equation. The model naturally prefers $^3P_2$ pairing because although the induced interaction at the Fermi surface is attractive for all values of the total angular momentum $J=0,1,2$, the sum $V^{\rm bare}+V^{\rm ind}$ is only attractive for $J=1,2$ and the net attraction is larger for $J=2$. In Fig.~\ref{fig:vector_gaps}, we show the $^3P_2$ pairing gaps calculated using the BCS formula in Eq.~\ref{eq:DPNR}. Results are shown for three choices of the coupling $G_V$  obtained by setting $F_0=2,3$ and $4$ at $n_B= 3~\nsat$. 

\begin{figure}[h]
\begin{center}
\includegraphics[width=0.85\textwidth]{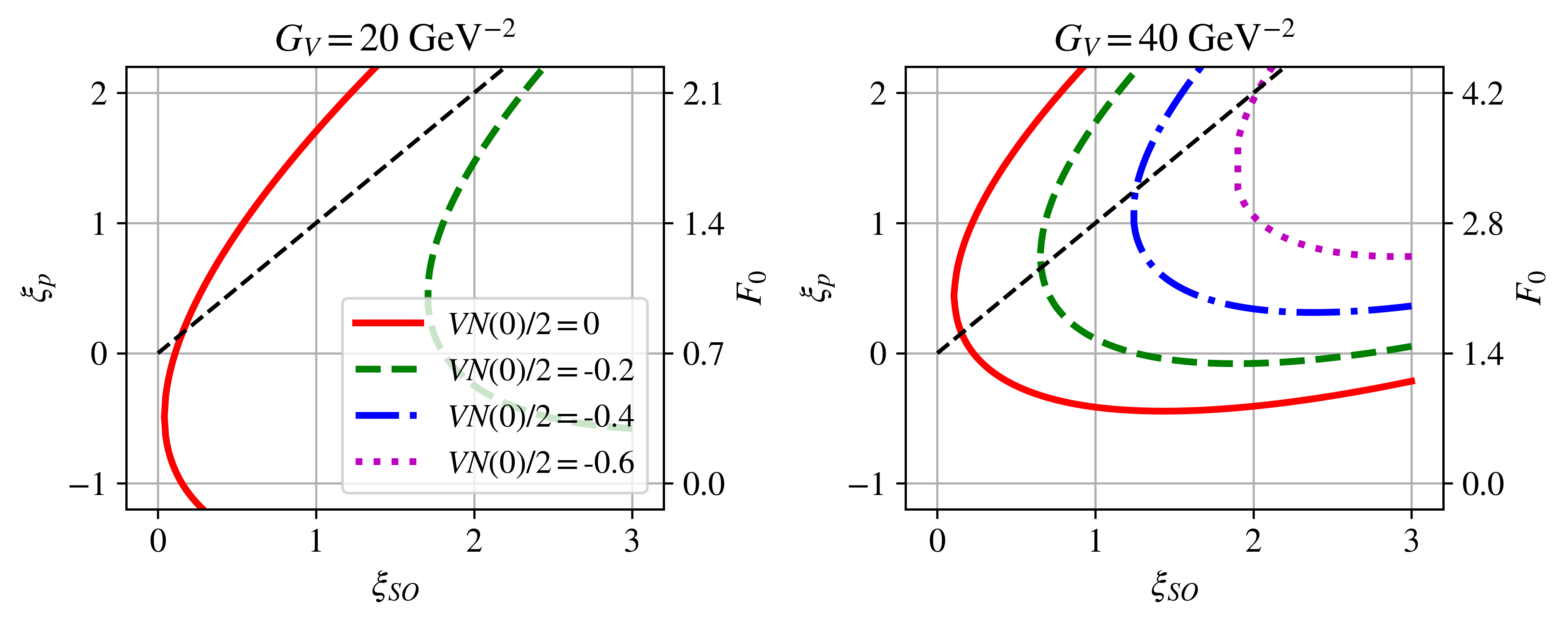}
\caption{Contours of constant $^3P_2$ potential in model B in the $\xi_p-\xi_{SO}$ plane for two choices of $G_V$ at $3\nsat$. $F_0$, calculated using Eq.~\ref{eq:F0G0} is also shown. The black dashed line shows where the bare interaction vanishes.}
\label{fig:vtot_2xi}
\end{center}
\end{figure}

To study the interplay between the central p-wave and the spin-orbit interactions, we consider scenario B, in which we introduce parameters $\xi_{\rm p}$ and  $\xi_{\rm SO}$ to control the strength of the central p-wave interaction and the spin-orbit interaction, respectively. In this case, we neglect the tensor coupling, and the LEC constants are given by:
\begin{equation}
    \bar{C}_0=G_V, \quad \bar{C}_2 = \frac{G_V}{\Lambda^2}, \quad \bar{C}'_2 = \xi_{\rm p} \frac{G_V}{\Lambda^2}  , \quad V_{SO} = - \xi_{SO} \frac{G_V}{\Lambda^2}, \quad V_T = 0
\end{equation}
As a generic choice at the same order as the nucleon and meson masses, we take $\Lambda = 1 \GeV$. Fig.\ref{fig:vtot_2xi} shows curves of constant $VN(0)/2$ in the space of $\xi_p$ and $\xi_{SO}$ for $G_V=20\GeV^{-2}$ and $G_V=40\GeV^{-2}$. Corresponding values of $F_0$ for each value of $\xi_p$ are shown on the right axis.  
The black line shows where the bare interaction is zero.  Above this line, the bare interaction is repulsive, and below it is attractive. The general behavior of the induced interaction is set by terms proportional to $\bar{C}^{\prime 2}_2$, $\bar{C}'_2 V_{SO}$, and $V_{SO}^2$ with some secondary effects from terms proportional to $\bar{C}_0 \bar{C}'_2$ and $\bar{C}_2 \bar{C}'_2$. Even though some of these terms enter at higher order in the gradient expansion, they are generally more important than lower-order terms in the induced interaction because they do not rely on the KL singularity to contribute to the $^3P_2$ potential. Of these five terms, the only ones that can be attractive are $\bar{C}_0 \bar{C}'_2$ and $\bar{C}_2 \bar{C}'_2$ when $\xi_p$ is positive, and $\bar{C}'_2 V_{SO}$ when $\xi_{SO}$ and $\xi_p$ are of the same sign. As a result, for stronger couplings, the overall interaction is repulsive when $\xi_p$ is negative and of reasonable size, even though that corresponds to a more attractive bare interaction. There is significant net attraction only when $\xi_p$ and $\xi_{SO}$ are both positive and $\xi_p$ is not much larger than $\xi_{SO}$ as this leads to a more repulsive bare interaction without producing enough induced attraction to match.

\begin{figure}[h]
\begin{center}
\includegraphics[width=0.7\textwidth]{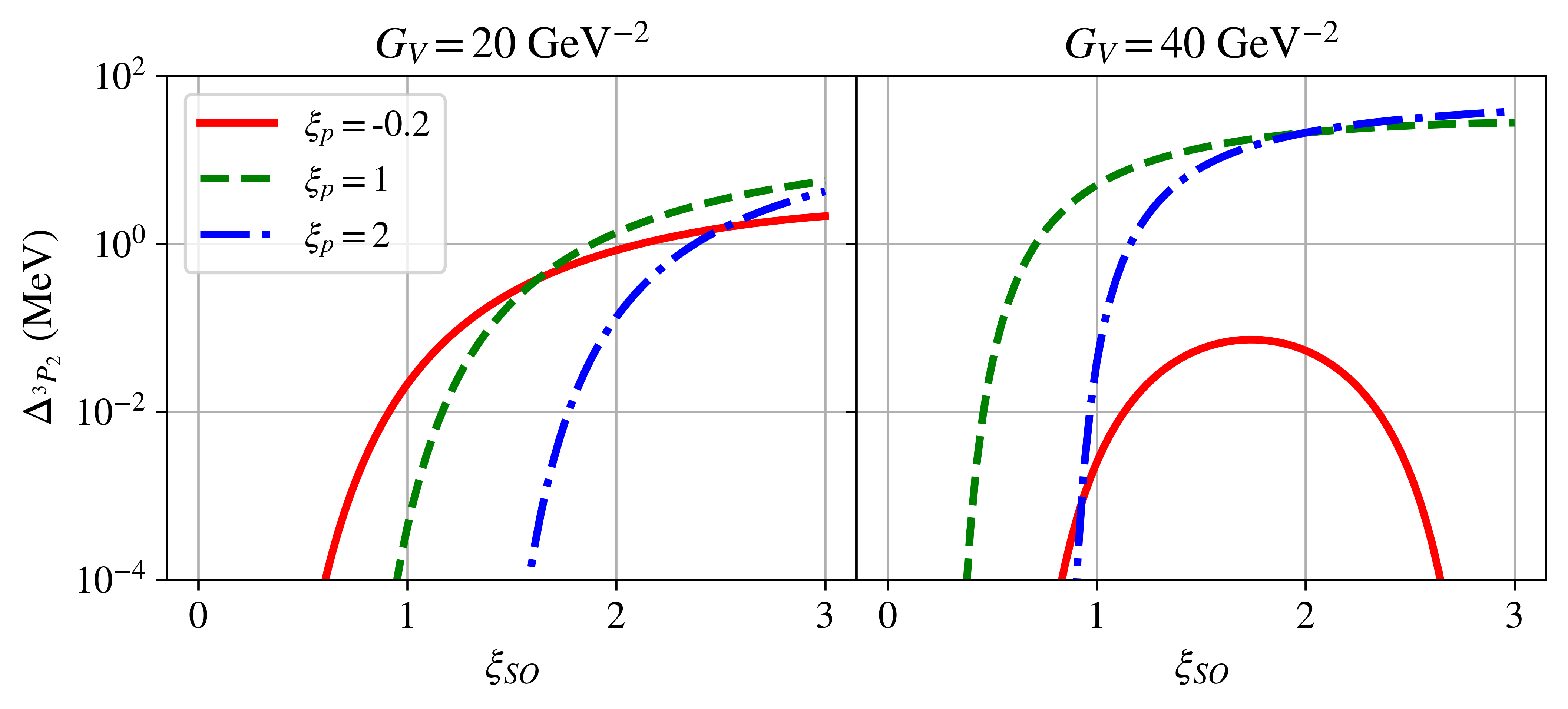}
\caption{$^3P_2$ gap as a function of $\xi_{SO}$ in model B for a few choices of $\xi_p$ and $G_V$ at $3\nsat$.}
\label{fig:gap_2xi}
\end{center}
\end{figure}

Fig.\ref{fig:gap_2xi} shows the $^3P_2$ gap as a function of $\xi_{SO}$ for a few choices of $\xi_p$ and the same choices of $G_V$. This interplay between the relative size of $\xi_p$ and $\xi_{SO}$ sensitively determines the size of the gap. This model is not detailed enough to make quantitative predictions, but the general trend remains that high-order terms in the expansion play an important role in determining the size of the gap.  An attractive spin-orbit appears to be necessary to have gaps of a reasonable size.  An attractive bare p-wave potential precludes pairing even though the bare interaction is stronger because of the repulsion due to the induced interaction.

To incorporate trends observed in the nucleon-nucleon phase shifts that have been measured up to $E_{\rm lab}\simeq 300$ MeV which correspond $p_{\rm CM}=\sqrt{m_n E_{\rm lab}/2}\simeq 375$ MeV, we consider scenario C in which we incorporate a non-zero $V_T$ by introducing a parameter $\xi_T$ that sets the strength of the tensor interaction, and $V_T = -G_V \xi_T / \Lambda^2$. This allows us to obtain any desired ordering of the p-wave phase shifts for $J=0,1,2$ and match scattering data that require a weakly attractive bare $^3P_2$ interaction and repulsive interactions in the $^3P_0$ and $^3P_1$ channels.  Since the induced interaction at ${\cal O}(m\kF^5)$ in Eq.~\ref{eq:order_kf5} neglected the part of the potential proportional to $\bar{C}'_2 V_T$, the results we present here are strictly only valid for $\xi_T \ll \xi_{SO}$. Thus, scenario C must be viewed as an initial exploration into the effects of $V_T$ to be continued in future work.

\begin{figure}[h]
\begin{center}
    \includegraphics[width=\textwidth]{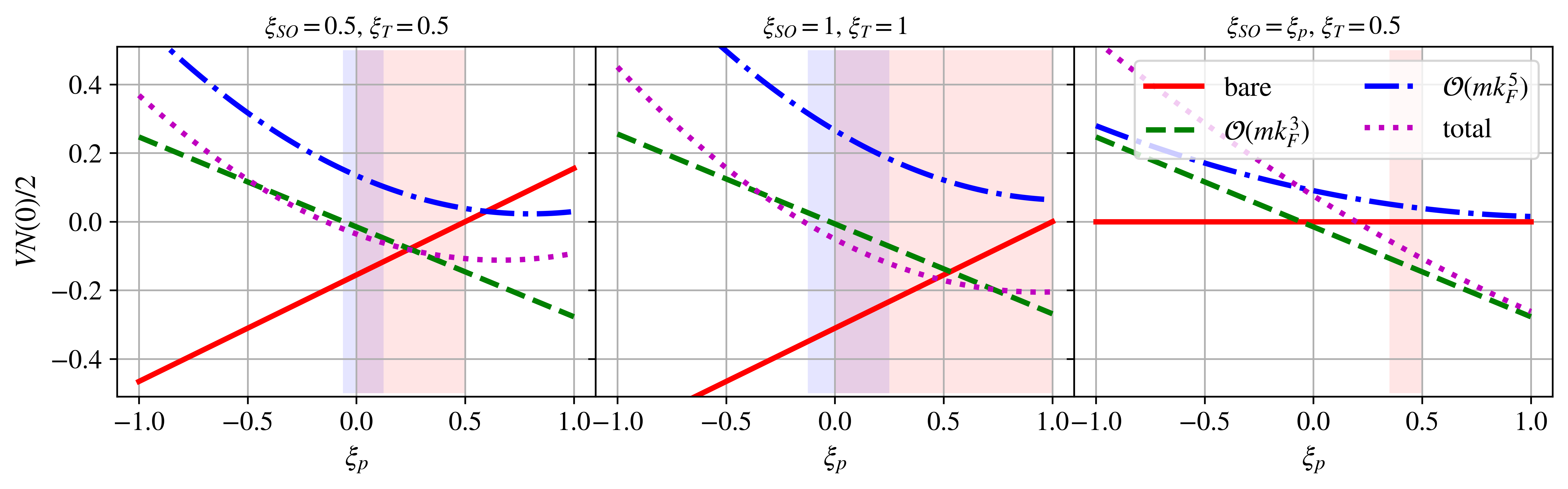}
    \includegraphics[width=\textwidth]{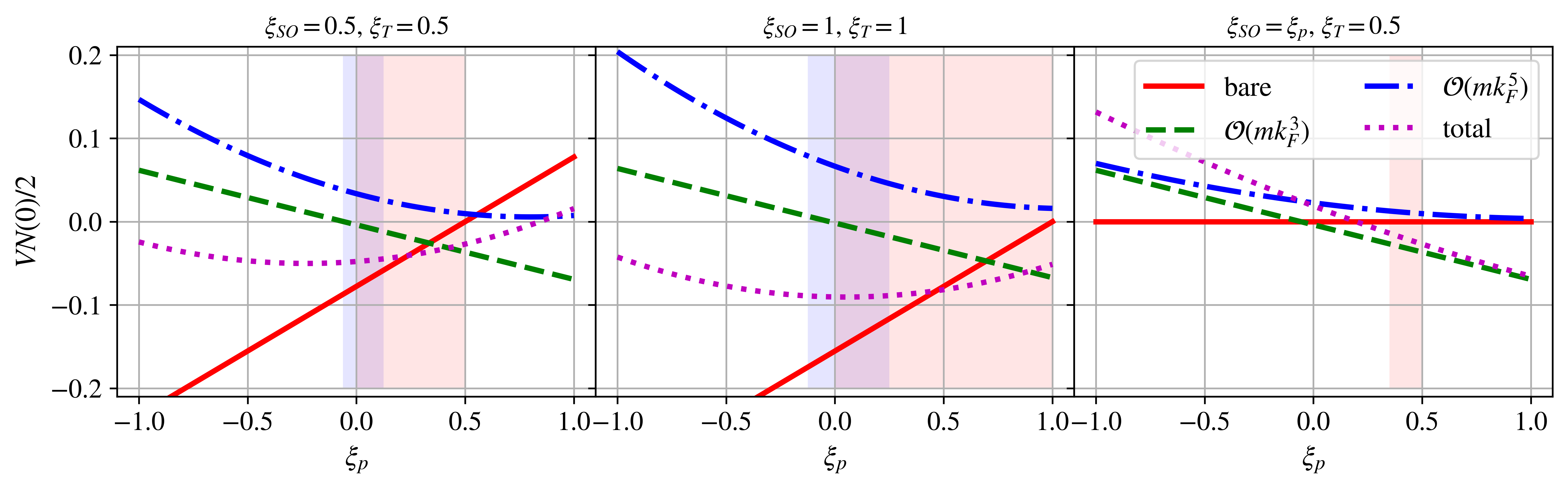}
    \caption{ The bare $^3P_2$ potential, the induced potential at $\mathcal{O}(mk_F^3)$, and the induced $V_{SO}$ at $n_B=3\nsat$. $G_V=40 \GeV^{-2}$ in the upper panel and $G_V=20\GeV^{-2}$ in the lower panel. Blue and orange bands indicate where $-3<\alpha < -1$ and $2<\beta<4$ respectively with the correct sign for all phase shifts for $\alpha$ and $\beta$ defined in Eq. \ref{eq:alpha_beta}.}
\label{fig:vbyorder_3xi}
\end{center}
\end{figure}

\begin{figure}[h]
\begin{center}
\includegraphics[width=\textwidth]{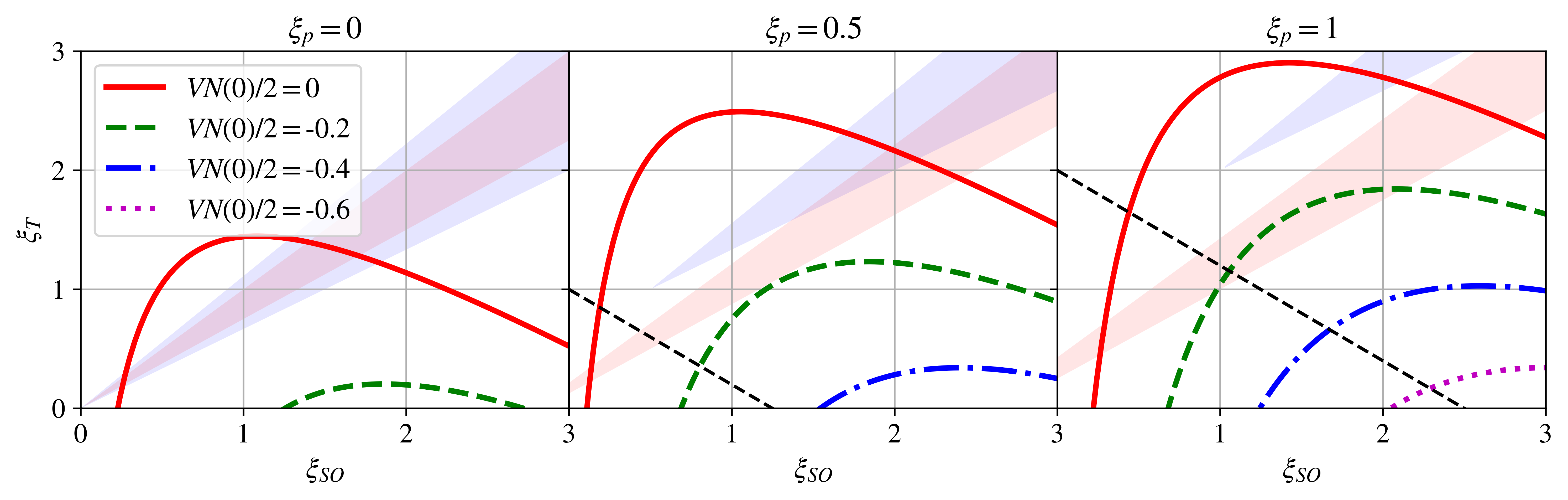}
\caption{Curves of constant total potential for different choices of $\xi_p$ at $3\nsat$. The black dashed line shows where the spin-orbit correction vanishes. The coupling constant is $G_V=40\GeV^{-2}$. Blue and orange bands indicate where $-3<\alpha < -1$ and $2<\beta<4$ respectively with the correct sign for all phase shifts for $\alpha$ and $\beta$ defined in Eq. \ref{eq:alpha_beta}.}
\label{fig:vindvso_3xi}
\end{center}
\end{figure}

To obtain the correct ordering of the p-wave interactions, the parameters must satisfy the following conditions.  To have an attractive bare $^3P_2$ potential, $\xi_{SO} > \xi_p$ and to have a repulsive bare $^3P_0$, $\xi_p + 2 \xi_{SO} - 3\xi_{T}/2 > 0$.  The condition $2\xi_{SO} / 5 < \xi_T < 2 \xi_{SO}$ ensures that the $^3P_1$ potential is most repulsive and $^3P_2$ is most attractive. We define $\alpha$ and $\beta$ to be the ratio between the bare potentials given by
\begin{equation}
\label{eq:alpha_beta}
    \begin{split}
        \alpha \equiv \frac{V_{^3P_2}}{V_{^3P_0}} = \frac{\xi_p - \xi_{SO}}{\xi_p + 2 \xi_{SO} - 3 \xi_T / 2} \\
        \beta \equiv \frac{V_{^3P_1}}{V_{^3P_0}} = \frac{\xi_p + \xi_{SO} + \xi_T}{\xi_p + 2 \xi_{SO} - 3 \xi_T/2}
    \end{split}
\end{equation}
Phase shifts for lab energies between $250 \MeV$ and $350 \MeV$ favor $\alpha$ between $-1$ and $-3$ and $\beta$ between $2$ and $4$. The blue and orange bands in Fig. \ref{fig:vbyorder_3xi} and Fig. \ref{fig:vindvso_3xi} identify regions $-3<\alpha<-1$ and $2<\beta<4$, respectively with the correct sign for each bare potential.  We only show regions with $\xi_{SO}>0$ and $\xi_T>0$ since this is required to obtain the correct ordering of p-waves phase shifts. These are also the signs favored by a tensor interaction arising from pion exchange and a spin-orbit force from heavy meson exchange.

To illustrate the relevance of the induced interaction, Fig. \ref{fig:vbyorder_3xi} shows the interaction for scenario C broken down into bare potential, $\mathcal{O}(mk_F^3)$ induced potential, and $\mathcal{O}(mk_F^5)$ corrections to the induced potential for a few choices of $\xi_{SO}$ and $\xi_T$ and the same two values of $G_V$ used for scenario B. In the right panel, $\xi_{SO}$ is fixed to match the value of $\xi_p$ so that the bare interaction is always zero, as is found in the meson exchange model. The range of $\xi_p$ chosen is motivated by the observation that meson exchange models predict $0.5 \lesssim \xi_p \lesssim 1.5$ and matching to phase shifts between 250 and 350 MeV predicts $-0.5 \lesssim \xi_p \lesssim 0$. As seen before, the $\mathcal{O}(mk_F^3)$ induced interaction is dominated by the term proportional to $\bar{C}_0 \bar{C}'_2$ and gives more attraction for positive $\xi_p$ and repulsion for negative $\xi_p$. The induced interaction does not exactly go through the origin for $\xi_p=0$, but the KL suppression of the terms that do not include $\bar{C}'_2$ or $V_{SO}$ is sufficient that it is not visible on this scale. Negative values of $\xi_p$ lead to more repulsive $\mathcal{O}(mk_F^5)$ corrections, with this effect being stronger for larger values of $\xi_{SO}$. 

Fig. \ref{fig:vindvso_3xi} shows the contours of constant potential for scenario C. As in scenario B, negative or zero $\xi_p$ corresponds to repulsion for most of the parameter space. However, unlike scenario B, when $\xi_T$ is of reasonable size, increasing $\xi_{SO}$ results in repulsion much more quickly. This is a result of the term proportional to $V_T V_{SO}$. The contribution of the spin-orbit terms can be easily summarized in terms of the constants of this model:
\begin{equation}
    V^{(5SO)}_{^3P_2} = \xi_{SO}\left( \xi_T + \frac{4\xi_{SO}}{5} - 2 \xi_p \right) \frac{G_V^2}{\Lambda^4} \frac{32mk_F^5}{945\pi^2}(43 + 24 \log 2)
\label{eq:vso_condition}
\end{equation}
In scenario A, $\xi_T$ is negative and $\xi_{SO} = \xi_p$, so the term in parentheses is always negative, and the spin-orbit corrections give additional attraction.  However, when we allow the constants to vary and take $\xi_T$ positive as is favored by pion exchange and phase shift data, much of the favored region of the phase diagram shows suppression due to these corrections.  The black dashed line in Fig. \ref{fig:vindvso_3xi} shows where $\xi_T + 4\xi_{SO}/5 - 2 \xi_p =0$.  The spin-orbit corrections suppress the potential above and to the right of this line, while the potential is enhanced below and to the left.  

\begin{figure}[h]
\begin{center}
    \includegraphics[width=0.7\textwidth]{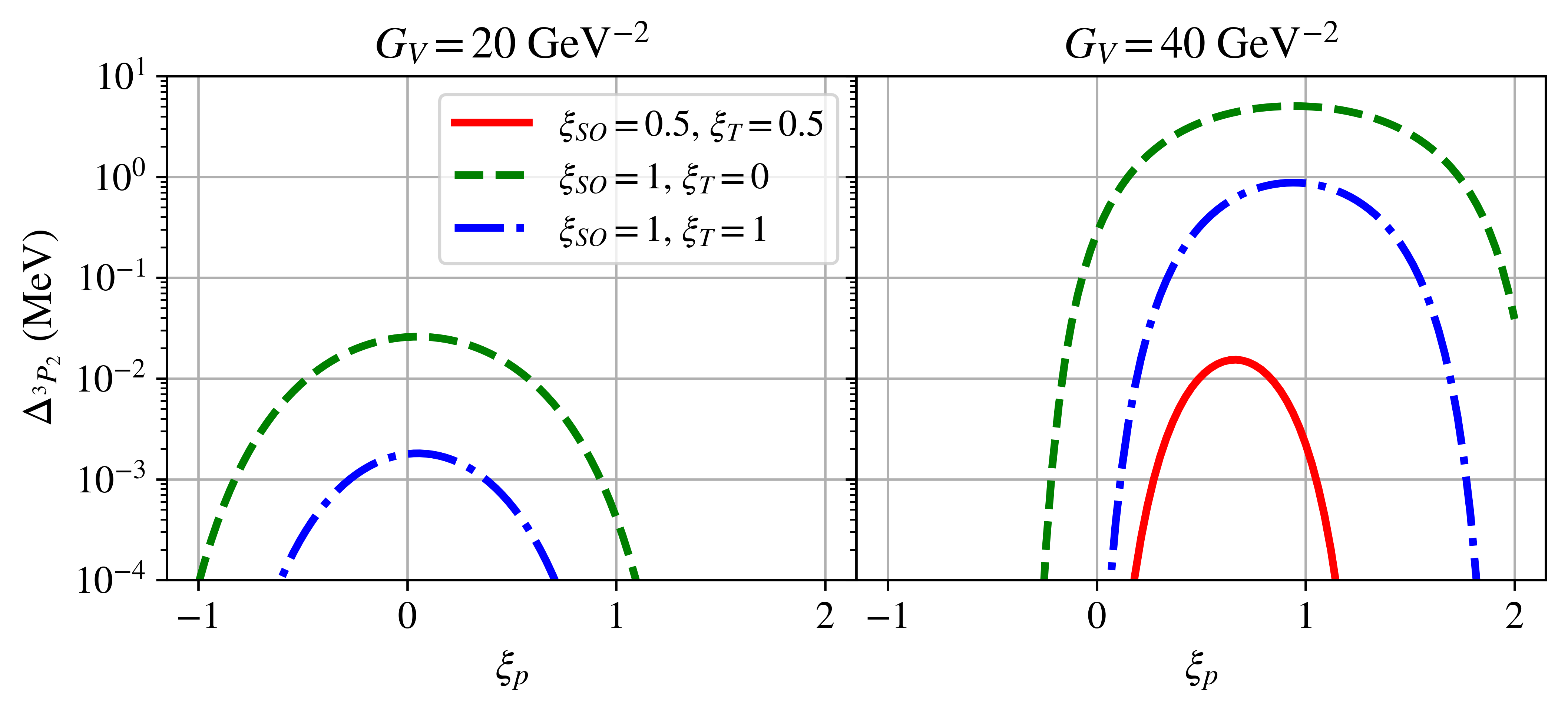}
    \caption{$^3P_2$ gap as a function of $\xi_p$ for a few choices of $G_V$, $\xi_{SO}$, and $\xi_T$ at $n_B=3\nsat$.}
\label{fig:gap_3xi}
\end{center}
\end{figure}

Fig. \ref{fig:gap_3xi} shows the $^3P_2$ gap at $3\nsat$ as a function of $\xi_p$ for a few choices of $\xi_{SO}$ and $\xi_T$.  For negative $\xi_p$, even though the bare interaction is more attractive, the induced interaction strongly suppresses the gap. The contribution of positive $\xi_T$ suppresses the gap. For positive $\xi_p$, MeV-scale gaps are possible. These results are sensitive to the value of $F_0$, as can be inferred by comparing the upper and lower panels of Fig.\ref{fig:vbyorder_3xi}. For smaller $F_0$, the relative importance of the induced interaction is diminished, and for most of the parameter space, the induced interaction suppresses an attractive bare interaction.  This still has a significant effect as the gap is exponentially sensitive to the potential. In this case, for almost all the explored parameter space, the induced interaction suppresses the gap by an order of magnitude or more.  

\section{Induced p-wave pairing in quark matter}
\label{sec:qmatter}

The inner cores of neutron stars may be made up of deconfined quark matter.  Quarks are expected to form a color superconductor at asymptotic densities, with the dominant pairing being in the color antitriplet channel (antisymmetric).  See \cite{Alford2001CSCReview} for a review. Color symmetric pairing is generally not considered since gluon exchange in the color sextet (symmetric) is repulsive, while the antitriplet channel is attractive.  At asymptotic densities, pairing all three colors and flavors (up, down, and strange) in color and flavor antisymmetric pairs is expected (the CFL phase). At lower densities where the strange quark mass cannot be neglected, up and down quarks can pair in the color antisymmetric channel. This pairing involves two colors, typically denoted as red and green, and is called the 2SC phase. In this phase, the strange quarks and blue-up and blue-down quarks are unpaired. 

Pairing of quarks of the same flavor and color was considered in Ref.~\cite{PRDAlfordNonLocking} in which possible pairing channels were found for strange quarks in color symmetric and antisymmetric channels for a bare interaction with the quantum numbers of gluon exchange. The attractive color symmetric channel they find is model-dependent and, in the case of massive quarks with all contributions from gluon exchange, receives competition from repulsive terms with the same spin and angular momentum quantum numbers but different chirality. Since they consider contact interactions without explicit derivative couplings, they explore $J=0$ and $1$.  Pairing of quarks of the same flavor was also studied in Ref.~\cite{Schmitt_spin_one_csc} in which attractive color antisymmetric channels with $J=1$ were considered, finding the transverse color-spin locked phase to be favored.

The possibility that quarks left over in the 2SC phase could pair due to the KL mechanism in QCD was first studied in Ref.~\cite{Schafer:2006ue}. This study investigated the KL effect in gauge theories where fermions interact via long-range forces that are dynamically screened due to Landau damping of the magnetic gauge bosons. Here, the energy dependence of the interaction plays a critical role, and the results of Ref.~\cite{Schafer:2006ue} indicate that a gap arises via a mechanism analogous to the Kohn-Luttinger effect but conclude that it is too small to be phenomenologically relevant. 

To assess if the KL mechanism could be relevant in quark matter with short-range interactions that are independent of energy, we shall calculate the induced interaction in the $^3P_2$ channel between quarks of the same flavor and color due to the short-range, flavor and color-independent, repulsive vector interaction. Such pairing would include the strange and the up and down blue quarks. In what follows, we shall focus on the induced interaction between strange quarks of the same color at moderate density when $m_s \gg k_{Fs}$  The specific question we address here is if repulsive short-range interactions introduced to stabilize quark matter inside neutron stars \cite{Baym_2018} can lead to pairing gaps of phenomenological relevance.

For concreteness, we consider a description of quark matter within the purview of the Nambu-Jona-Lasino(NJL) models (see \cite{Buballa:2005} for a comprehensive review). In these models,  defined by the interaction Lagrangian \cite{Buballa:2005,Baym_2018,PRDSongBaymGluonRepulsion}  
\begin{equation}
\label{eq:NJL}
{\cal L}_{V} = G(\bar{q} q)^2 + H  (\bar{q}\bar{q}) (qq) - g_V ~(\bar{q} \gamma_\mu q)^2\,, 
\end{equation}
where $G$ and $H$ are the four-fermion scalar quark-antiquark and diquark coupling strengths, and vector coupling $g_V$ is introduced to generate higher pressures as noted earlier. The scalar interaction between quarks and antiquarks leads to a non-trivial vacuum with $\langle \bar{q} q\rangle \neq 0$ that spontaneously breaks chiral symmetry and the coefficient $G$ is determined by hadron masses and the pion decay constant in the vacuum. For typical momentum cut-off $\Lambda_{\rm NJL} \approx  600$ MeV,  $G \Lambda_{\rm NJL}^2 \simeq 2$ \cite{Buballa:2005}. At densities of interest to neutron stars, chiral symmetry remains broken. The constituent strange quark mass is expected to be $300-500$ MeV, while the up and down quark masses can be significantly smaller.  The diquark coupling $H$ and the vector coupling $g_V$ are expected to be of similar size because they can be thought of as arising from the same underlying high-energy color current-current interactions in QCD \cite{PRDSongBaymGluonRepulsion}. Their values at the densities of interest to neutron stars are determined phenomenologically. The diquark coupling $H$, which encodes the attraction in the color antisymmetric channel, leads to s-wave pairing between quarks. For $H \simeq G$, the s-wave pairing gap between up and down quarks is about $50$ MeV and is typically inadequate to induce pairing between strange quarks and light quarks \cite{Alford:2007xm}, as mentioned above. The analysis of the quark matter EOS in \cite{Baym_2018,Baym_2019} concluded that vector coupling needed to be of moderate size with $g_V \simeq G$ to support the large sound speed needed to support a two solar mass neutron star. 

First, we note that the contribution to the induced interaction from the closed fermion loop (the first diagram in Fig.~\ref{fig:kldiagrams}) is enhanced by a factor $N_f N_c$. This is because the bare vector interaction introduced to stiffen the quark matter EOS is independent of color and flavor. Thus, in contrast to the one-component Fermi system,  where the contribution from the closed fermion loop was canceled by the diagram that encoded the vertex corrections, in quark matter with $N_f = N_c = 3$, the first diagram in Fig.~\ref{fig:kldiagrams} makes the dominant contribution to the induced potential.  In computing this diagram, the up and down quarks must be treated as relativistic particles, leading to a somewhat more complicated expression.  After doing the Matsubara sum and noting that $\bar{u}_3 \slashed{q} u_1 = \bar{u}_4 \slashed{q} u_2$ for $u_1$ and $u_2$ incoming and $\bar{u}_3$ and $\bar{u}_4$ outgoing spinors, the induced potential from the first diagram is given by:
\begin{equation}
    \label{eq:v_ind_quark}
    V^{\rm ind} = g_V^2 (\bar{u}_3 \gamma_\mu u_1)  (\bar{u}_4 \gamma_\nu u_2) \left( \frac{E_k + m_s}{2m_s} \right)^2 \sum_{f,c} \int \frac{\ell  d\ell d\Omega_\ell}{4\pi^3 q E_\ell} \frac{\Theta(k_{fc} - \ell)}{c_{q\ell} - q / 2\ell} (2 \ell^\mu \ell^\nu - g^{\mu \nu} \vec{\ell} \cdot \vec{q})
\end{equation}

\begin{figure}[h]
\begin{center}
\includegraphics[width=0.5\textwidth]{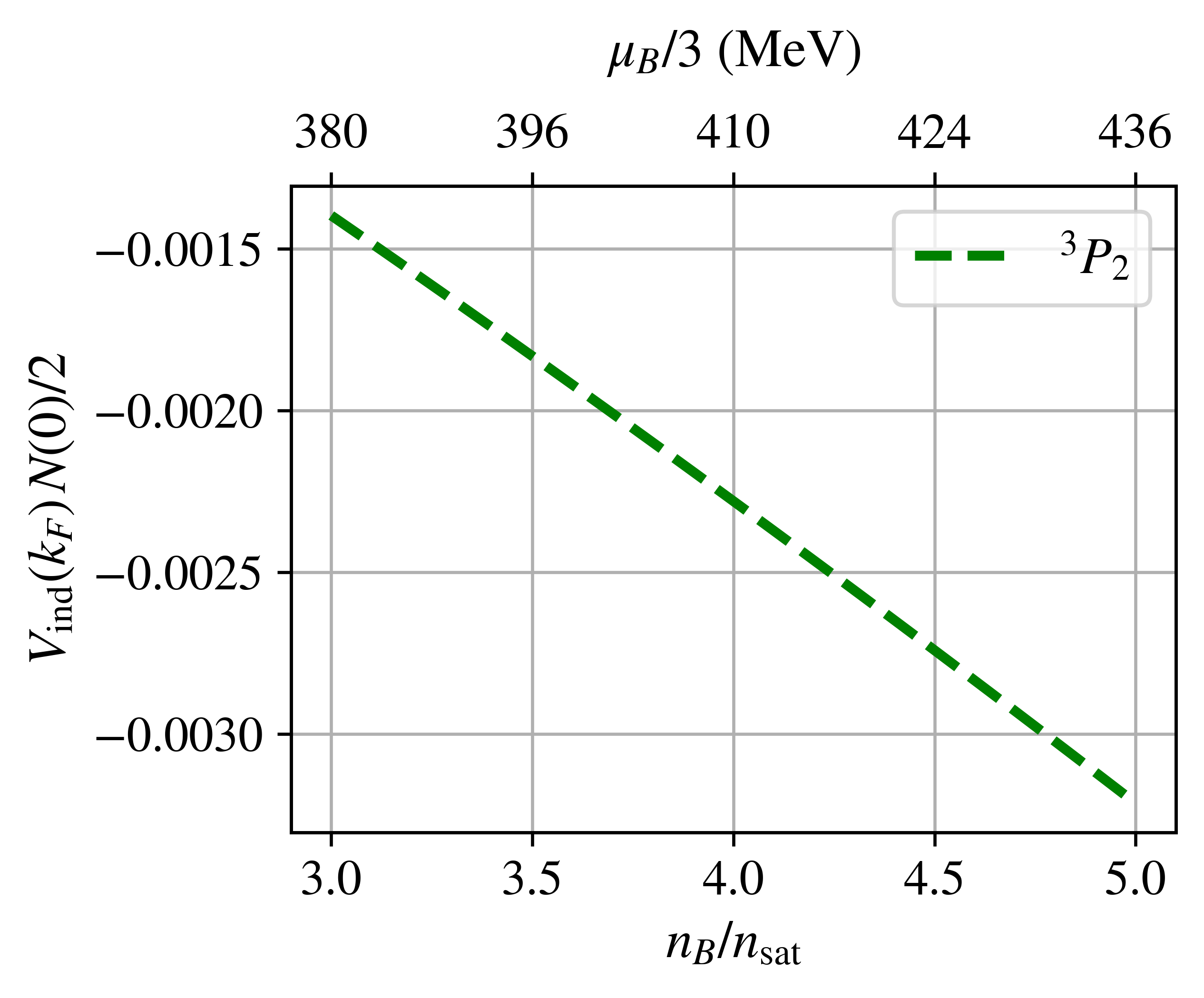}
\caption{Induced p-wave potential between strange quarks for $g_V = 2/\Lambda^2_{NJL}$, $\Lambda_{NJL} = 600 \MeV$, and $m_s = 350 \MeV$. }
\label{fig:quark_potential}
\end{center}
\end{figure}

The calculation of the induced interaction, including both the electric ($\bar{u} \gamma^i u$ for $i = 0$) and magnetic ($\bar{u} \gamma^i u$ for $i = (1, 2,3)$) components is unwieldy. In what follows, we shall focus on the electric component as the magnetic component is suppressed by the strange quark mass. In this case, setting $\mu=\nu=0$ in Eq.~\ref{eq:v_ind_quark} we find that 
\begin{equation}
    V^{\rm ind} =  \frac{g_V^2 \delta_{13} \delta_{24}}{2\pi^2 q} \sum_c \int \frac{d\ell dc_{q\ell}}{c_{q\ell} - q / 2\ell} \left[ \sum_{f=u,d} ( 2\ell^2 - \ell q c_{q\ell}) \Theta(k_{fc}-\ell)  + 2\ell m_s \Theta(k_{sc}-\ell) \right]\,. 
\end{equation}
After performing the momentum integrals, we find that   
\begin{equation}
    V^{\rm ind} = g_V^2 \delta_{13} \delta_{24} \sum_c \left[ \sum_{f=u,d} \left( -\frac{k_{fc}^2}{2\pi^2} +\frac{q^2}{2} U_0^{\rm rel} \left( \frac{q}{k_{fc}} \right) - 2k_{fc}^2 U_2^{\rm rel} \left( \frac{q}{k_{fc}} \right) \right) - 2 U(q) \right]
\end{equation}
where the relativistic Lindhard functions $U_0^{\rm rel}$ and $U_2^{\rm rel}$ defined in Appendix \ref{App:QuarkInduced}. 

Note that by not including the magnetic part of the vector integration ($\bar{u} \gamma^i u$ for $i = (1, 2,3)$), explicit dependence on $J$ has been removed, and all the p-waves have the same potential. Nonetheless, the $^3P_2$ gap remains of primary interest because the bare interaction vanishes for $J=2$, while it is repulsive for $J=0,1$. Fig. \ref{fig:quark_potential} shows the induced p-wave potential for strange quarks of mass $350$ MeV, for $g_V=2/\Lambda_{NJL}^2$ and $\Lambda_{NJL}=600\MeV$. Densities of each color and flavor are determined assuming 2SC pairing of up and down quarks, charge neutrality, and beta equilibrium. Since the pairing gap $\Delta \simeq \mu~ \exp{\left(2/V_{\rm ind}(k_F) N(0)\right)}$, from Fig.~\ref{fig:quark_potential} we can deduce that the induced potential is too small to be relevant for neutron star phenomenology. This is mostly due to the fact that the value of $g_V$ needed in NJL models to support massive neutron stars is significantly smaller than the coupling we considered for nucleons. 

\section{Implications for Neutron Stars}
\label{sec:implications}

The thermal evolution of neutron stars, especially those that are reheated by accretion from a companion at late times, is sensitive to heat capacity and neutrino emissivity in their cores \cite{Yakovlev:2004iq,Page:2009,Potekhin:2015qsa}. The neutrino emissivity and the specific heat of dense matter are both strongly modified by Cooper pairing. When the pairing gap is large compared to the temperature, the neutrino emissivity and the specific heat are exponentially suppressed by the factor $\exp{(-\Delta/ k_B T)}$.  Additionally, in the vicinity of the critical temperature, Cooper pair breaking and formation (PBF) processes enhance the neutrino emissivity. This enhancement is especially important for neutron Cooper pairing in the $^3P_2$ channel in the core of the neutron star \cite{Page:2009,Potekhin:2015qsa}. Studies of isolated neutron star cooling reported in Ref.~\cite{Page:2009} that include the modified URCA ($nn\rightarrow np e^- \bar{\nu}_e$ and $e^- p n\rightarrow n n  \nu_e$) reactions and the PBF process but discount the possibility of other more rapid neutrino emission processes such as direct URCA \cite{Lattimer:1991ib} find that a critical temperature for $^3P_2$ pairing, $T_c\approx \Delta_{^3P_2}/1.7$, that is larger than $5 \times  10^8$ K ($\approx 50$ keV) throughout the inner core would be disfavored by observations. This favors a scenario in which the $\Delta_{^3P_2}$ is suppressed at the modest density encountered in the outer core due to the competition between the interactions induced by the central and spin-orbit components of the nuclear forces \cite{Schwenk:2003bc} but is insensitive to the behavior of the gap at higher density. 

Accreting neutron stars exhibit a diversity of cooling behaviors, and a few neutron stars show behavior that requires rapid neutrino cooling \cite{Yakovlev:2004iq,Brown:2017gxd}. Such rapid neutrino cooling can be realized in the dense nuclear matter when the proton fraction in the core exceeds about $11\%$ to lift kinematic restrictions on the direct URCA reactions $e^- + p \rightarrow n + \nu_e$ and $n \rightarrow e^- + p + \bar{\nu}_e$ \cite{Lattimer:1991ib}. In addition, rapid cooling would also require $^3P_2$ pairing to be absent at high density. Our finding that the induced interaction disfavors $^3P_2$ when the spin-orbit and tensor forces are strong and attractive provides some insight into the conditions necessary to realize unpaired neutron matter at high density characterized by a high sound speed. On the other hand, if the central component of the p-wave interaction is strongly repulsive and the non-central components are weak, the induced interaction favors $^3P_2$ pairing between neutrons, and rapid neutrino cooling cannot be realized in nuclear matter at high density. In this scenario, rapid cooling in neutron stars would require new ungapped fermionic excitations, such as hyperons or quarks, to enable the direct URCA reaction.   

\begin{figure}[h]
\begin{center}
\includegraphics[width=0.5\textwidth]{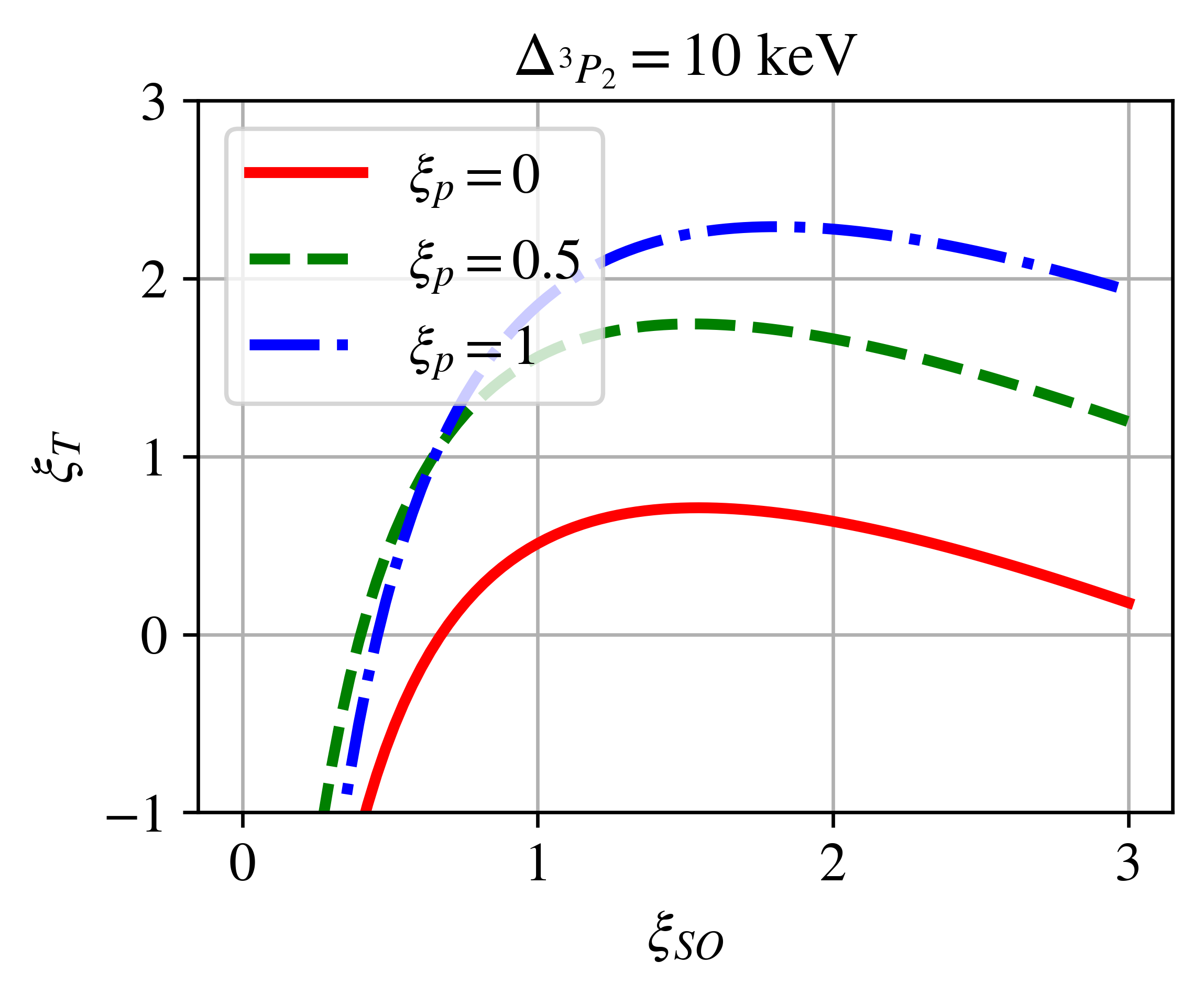}
\caption{Contours for $\Delta_{^3P_2} = 10 \keV$ for $G_V=40 \GeV^{-2}$ at $3\nsat$ for a few choices of $\xi_p$. Below and to the right of the contour, the gap is larger than 10 keV.}
\label{fig:all xi 3d contour}
\end{center}
\end{figure}
In transiently accreting neutron stars, inference of the heat deposition due to deep crustal heating from observations of accretion outbursts and the inference of the core temperature from subsequent observation in quiescence have been used to derive a lower limit to the neutron star core heat capacity \cite{Cumming:2016weq,Brown:2017gxd}. Further, if the neutron stars cooling can be observed during quiescence, an upper limit on the core heat capacity can also be deduced from observations \cite{Cumming:2016weq,Brown:2017gxd}. For neutron stars in the low mass x-ray binaries KS 1731-260, MXB 1659-29, and XTE J1701-462, with core temperatures in the range $10^7-10^8$ K, the lower limit was found to be a factor of a few below the core heat capacity expected if neutrons and protons in the core are paired. However, upper limits from future cooling observations in these systems could constrain the extent of neutron pairing in the neutron star core. For example, the analysis in Ref.~\cite{Brown:2017gxd} suggests that if the neutron star in MXB 1659-29 cools by about 4\% during a 10-year period, a very large fraction of the neutrons in the neutron star core must be superfluid with a gap that is much larger than a few keV. If observed, it would disfavor a large attractive tensor interaction and would require an attractive spin-orbit interaction as shown in Fig.\ref{fig:all xi 3d contour}. Repulsive bare p-wave interactions permit larger regions of parameter space, while attractive bare p-wave interactions are more restrictive.

\section{Conclusion}
\label{sec:Conclusions}
We have calculated the induced potential between fermions at the Fermi surface to study the role of polarization effects in the dense medium. We find that short-range repulsive interactions due to the exchange of heavy vector mesons between neutrons, whose strength is related to the Fermi liquid parameter $F_0$ and the sound speed at high density, induce an attractive p-wave potential. Using a model that allows us to independently vary the strength of central and non-central p-wave interactions, we have investigated the competition between the bare and induced interactions to determine the conditions necessary to realize $^3P_2$ pairing in neutron matter at high density. When neutron matter is characterized by a large speed of sound $c_s^2 > 1/3$ and $F_0 \gtrsim 2$, the induced interaction plays an important role. We find that 
\begin{itemize}
\item The contribution to the induced interaction in a particular partial wave arising from terms in the bare interaction that do not contribute to that partial wave are suppressed because their contribution is strongly influenced by the KL singularity at $q = 2\kF$.  For this reason, the bare p-wave and the spin-orbit interactions are generally more important than lower-order terms for the induced $^3P_2$ potential.
\item The induced interaction favors $^3P_2$ pairing if the central component of the s-wave and p-wave interaction are strongly repulsive and the non-central components are small. The resulting gap, $\Delta_{^3P_2}$, can be in the range $0.1- 10$ MeV in the neutron star core and is exponentially sensitive to the induced potential. 
\item When the central p-wave and the spin-orbit interaction are both strong and attractive, the induced interaction is repulsive. Although the bare interaction is strongly attractive, the induced repulsion can preclude pairing or suppress $\Delta_{^3P_2}$ by orders of magnitude. 
\item In the presence of a strongly attractive spin-orbit interaction, the induced interaction favors $^3P_2$ pairing when the central p-wave is repulsive. Pairing persists even when the strength of the central p-wave repulsion is greater than the attractive spin-orbit interaction.  
\end{itemize}
An important caveat to these findings is our assumption that the bare interaction at the Fermi surface is well represented by Eq. \ref{eq:vnnso}. Further, at the high momenta of relevance when $n_B > 2 \nsat$, the nucleon-nucleon potential, and thereby the parameters of our model, are not well constrained by scattering data. Nonetheless, results obtained within the purview of the model allowed us to explore the connection between pairing and the strong repulsive central interactions needed to generate a high sound speed and large $F_0$ at densities expected in the cores of massive neutron stars. Our calculation, which includes the effect due to strong spin-orbit forces, provides useful formulae to gauge the interplay between repulsive central interaction and attractive spin-orbit interactions. However, further study of the role of strong tensor interactions warrants further study. 

Another aspect that warrants mention is the role of many-body forces. Although we have not explicitly accounted for them in our study here, earlier work has demonstrated that three-body forces can be incorporated through a density-dependent two-body potential that can then be constructed by normal ordering the three-body force with respect to a convenient reference state, such as the ground state of the non-interacting many-body system \cite{Hebeler:2009iv,Holt:2019bah}. Including the three-body force would thereby introduce a density dependence to the parameters of our model that set the strength of the two-body s-wave and p-wave interactions in dense matter. We believe the large range of parameter values we explored should be sufficient to account for corrections due to many-body forces partially. The density-dependence of the two-nucleon partial-wave matrix elements at the Fermi surface and the correlation between the parameters induced by the 3-body forces will be explored in future work. Finally, as cautioned earlier, the magnitude of the $^3P_2$ gap was calculated using the BCS approximation, which restricted the interaction to the Fermi surface. More work is needed to assess the reliability of this approximation.  

\section{Acknowledgements}
The U.S. DOE supported the work of M. K. and S. R. under Grant No. DE-FG02- 00ER41132.  We thank Silas Beane, Roland Farrell, Bengt Friman, Yuki Fujimoto, and Achim Schwenk for their suggestions and helpful discussions. S. R. also thanks the members of the N3AS Physics Frontier Center, funded by the NSF Grant No. PHY-2020275 for useful conversations.    
\appendix{}
\section{Induced interactions}
\label{App:InducedInteraction}
In this appendix, we derive analytic results for the induced interaction in two steps. First, for the sake of simplicity and clarity, we assume that the bare potential only contains a momentum-independent s-wave interaction characterized by the $C_0$ and $\tilde{C}_0$, and spin-orbit force with strength $V_{SO}$. In this case, the ZS diagram involves the product $V_L \times V_R$, where 
\begin{equation}
    \begin{split}
        V_L &= C_0 (\delta_{13} \delta_{ab} - \delta_{1b} \delta_{a3}) + \tilde{C}_0 (\sigma_{13} \cdot \sigma_{ab} - \sigma_{1b} \cdot \sigma_{a3}) 
        - V_{SO} 2 i q \times (\ell +k') \cdot (\sigma_{13} \delta_{ab} + \sigma_{ab} \delta_{13})\\
        V_R &= C_0(\delta_{24} \delta_{ba} - \delta_{2a} \delta_{b4}) + \tilde{C}_0(\sigma_{24} \cdot \sigma_{ba} - \sigma_{2a} \cdot \sigma_{b4}) 
        + V_{SO} 2 i q \times (\ell - k) \cdot (\sigma_{24} \delta_{ba} + \sigma_{ba} \delta_{24})\,.
    \end{split}
\end{equation}
Evaluating term-by-term, we find that the $C_0^2$ contribution is given by 
\begin{equation}
\begin{split}
    C_0^2 \sum_{ab = \{\uparrow, \downarrow\}} (\delta_{13} \delta_{ab} - \delta_{1b} \delta_{a3})(\delta_{24} \delta_{ba} - \delta_{2a} \delta_{b4}) 
    = C_0^2 \delta_{14} \delta_{23}
\end{split}
\end{equation}
To calculate the $\tilde{C}^2_0$ contribution, we use the following identities:
\begin{equation}
\label{eq:spinid} 
    \begin{split}
        \sum_{ab = \{\uparrow, \downarrow\} } \sigma_{ab}^i \sigma_{ba}^j &= 2 \delta^{ij}  \\
        \sum_{b = \{\uparrow, \downarrow \} } \sigma_{bc}^j \sigma_{ab}^i  &= \chi^\dagger_b \sigma^j \sigma^i \chi_a = \delta_{ab} \delta^{ij} - i \varepsilon^{ijk} \sigma_{ab}^k \\
        \sum_{bc=\{\uparrow, \downarrow \} } \sum_{i = 1}^3 \sigma_{cd}^i \sigma_{bc}^j \sigma_{ab}^i &= \sum_{i=1}^3 \chi_d^\dagger \sigma^i (2\delta^{ij} - \sigma^i \sigma^j) \chi_a = -\sigma_{ad}^j
    \end{split}
\end{equation}
to find that 
\begin{equation}
    \begin{split}
        \tilde{C}^2_0 \sum_{ab = \{\uparrow, \downarrow\}} (\sigma_{13} \cdot \sigma_{ab} - \sigma_{1b} \cdot \sigma_{a3}) (\sigma_{24} \cdot \sigma_{ba} - \sigma_{2a} \cdot \sigma_{b4}) 
        = \tilde{C}_0^2 (4 \sigma_{13} \cdot \sigma_{24} + 3 \delta_{14} \delta_{23} + 2 \sigma_{14} \cdot \sigma_{23})
    \end{split}
\end{equation}
The $C_0 \tilde{C}_0$ contribution is calculated by noting that $\sum_{ab} \delta_{ab} \sigma_{ba} = \mbox{Tr} [ \sigma] = 0$ and $\sum_b \sigma_{bc} \cdot \sigma_{ab} = 3 \delta_{ac}$. Explicitly,  
\begin{equation}
    \begin{split}
        C_0 \tilde{C}_0 &\sum_{ab = \{\uparrow, \downarrow\} } [(\delta_{13} \delta_{ab} - \delta_{1b} \delta_{a3})(\sigma_{24} \cdot \sigma_{ba} - \sigma_{2a} \cdot \sigma_{b4}) 
        + (\sigma_{13} \cdot \sigma_{ab} - \sigma_{1b} \cdot \sigma_{a3})(\delta_{24} \delta_{ba} - \delta_{2a} \delta_{b4})] \\
        &= C_0 \tilde{C}_0[-2(3\delta_{13} \delta_{24} + \sigma_{13} \cdot \sigma_{24}) + 2 \sigma_{14} \cdot \sigma_{23}]
    \end{split}
\end{equation}
We have calculated the leading order contributions from the spin-orbit interaction, proportional to $C_0~V_{SO}$ and $\tilde{C}_0~V_{SO}$ and find that their contributions vanish. First, consider the $C_0~V_{SO}$ term  
\begin{equation}
\label{eq:vovso1}
\begin{split}
    C_0 V_{SO} &\sum_{ab = \{\uparrow, \downarrow\} } [(\delta_{13} \delta_{ab} - \delta_{1b} \delta_{a3} ) 2i q \times (\ell - k) \cdot (\sigma_{24} \delta_{ba} + \sigma_{ba} \delta_{24}) \\
    &+ (\delta_{24} \delta_{ba} - \delta_{2a} \delta_{b4})2i q \times (-\ell -k') \cdot (\sigma_{13} \delta_{ab} + \sigma_{ab} \delta_{13})] \\
    &= 2i C_0 V_{SO} [q \times (\ell -k) \cdot (2\delta_{13} \sigma_{24} - \sigma_{24} \delta_{13} - \sigma_{13} \delta_{24}) \\
    &+ q \times (-\ell -k') \cdot (2 \delta_{24} \sigma_{13} - \sigma_{13} \delta_{24} - \sigma_{24} \delta_{13})]\,.
\end{split}
\end{equation}
Eq.~\ref{eq:vovso1} can be simplified further by noting that terms proportional to $q \times \ell$ vanish upon integrating over the angle $\theta_{q\ell}$ and using the fact that $q \times k = q \times k' = -q \times q' /2$. We find the induced interaction proportional to $C_0~V_{SO}$  
\begin{equation}
    \begin{split}
        i C_0 V_{SO}[q \times q' \cdot (2\delta_{13} \sigma_{24} - \sigma_{24} \delta_{13} - \sigma_{13} \delta_{24}+2\delta_{24} \sigma_{13} - \sigma_{13} \delta_{24} - \sigma_{24} \delta_{13})] = 0
    \end{split}
\end{equation}

To see that $q \times \ell$ terms vanish, notice that the only angular dependence from the loop integral is on the angle between $q$ and $\ell$.  Consider the integral $\int d\Omega_\ell \hat{\ell} \cdot \hat{u} f(\hat{\ell} \cdot \hat{q})$ where $f(\hat{\ell} \cdot \hat{q})$ contains the angular dependence of the loop integral and $\hat{\ell} \cdot \hat{u}$ corresponds to terms like $q \times \ell \cdot \sigma$.  Rotate $\Omega_\ell$ so that $\hat{q}=\hat{z}$ and $\phi_\ell=0$ corresponds to the azimuthal angle of $\hat{u}$ calling these angles $\theta_{q\ell}$ and $\phi_{u\ell}$.  Also define the polar angle of $\hat{u}$ as $\theta_{uq}$  Now $\hat{\ell} \cdot \hat{u} = \sin{\theta}_{q \ell} \cos{\phi}_{u\ell} \sin{\theta}_{uq} + \cos{\theta}_{q \ell} \cos{\theta}_{uq}$.  Doing the integral $\int d\phi_{u\ell} \cos{\phi}_{u\ell}=0$ so the only term that survives is proportional to $\cos{\theta}_{uq}$.  In the spin-orbit terms, $\ell$ always enters as $q \times \ell \cdot \sigma = \ell \cdot (\sigma \times q)$ with $\sigma \times q$ orthogonal to $q$, so this contribution always vanishes.

Similarly, the contribution proportional to $\tilde{C}_0 V_{SO}$ can also be simplified by making the substitutions $2iq \times (\ell -k) \rightarrow iq \times q'$ and $2iq \times (-\ell -k') \rightarrow iq \times q'$ and using the identities in Eq.~\ref{eq:spinid}. We find that 
\begin{equation}
\begin{split}
    \tilde{C}_0 V_{SO} &\sum_{ab = \{\uparrow, \downarrow\} } [(\sigma_{13} \cdot \sigma_{ab} - \sigma_{1b} \cdot \sigma_{a3}) i q \times q' \cdot (\sigma_{24} \delta_{ba} + \sigma_{ba} \delta_{24}) \\
    &+(\sigma_{24} \cdot \sigma_{ba} - \sigma_{2a} \cdot \sigma_{b4}) iq \times q' \cdot (\sigma_{13} \delta_{ab} + \sigma_{ab} \delta_{13})] \\
    &= \tilde{C}_0 V_{SO} ~iq \times q' \cdot (2\sigma_{13} \delta_{24} -3 \sigma_{24} \delta_{13} + \sigma_{13} \delta_{24} \\
    &+ 2\sigma_{24} \delta_{13} - 3 \sigma_{13} \delta_{24} + \sigma_{24} \delta_{13}) \\
    &= 0\,.
\end{split}
\end{equation}
Thus, spin-orbit terms do not contribute to the induced interaction at leading order in $V_{SO}$. Up to this order, including all of the non-zero terms associated with the product $V_L\times V_R$ and performing the particle-hole loop integration, we find that the induced interaction due to the ZS diagram is given by 
\begin{equation}
\begin{split}
    V^{\rm ind}_{ZS}&= - U(q) \left[ (C_0^2+3\tilde{C}_0^2)\delta_{14} \delta_{23} - 6C_0 \tilde{C}_0 \delta_{13} \delta_{24})  \right]  \\
     &- U(q) \left[(4\tilde{C}_0^2 - 2C_0 \tilde{C}_0) \sigma_{13} \cdot \sigma_{24} + (2\tilde{C}_0^2 + 2C_0 \tilde{C}_0)\sigma_{14} \cdot \sigma_{23}) \right]
\end{split}
\end{equation}
where 
\begin{equation}
\begin{split}
    U(q) &= -\frac{1}{\beta} \sum_{\ell_0} \int \frac{d^3 \ell}{(2\pi)^3} \frac{1}{\ell_0 - \ell^2/2m} \frac{1}{\ell_0 - (\ell+q)^2/2m} \\
    &= -\frac{m}{2\pi^2 q} \int_0^{k_F} \ell d\ell \int_{-1}^1 \frac{d\cos{\theta}_{q\ell}}{\cos{\theta}_{q\ell} - q/2\ell} \\
    &= -\frac{mk_F^2}{2\pi^2 q} \left[ -\frac{q}{2k_F} + \frac{1}{2} \left( 1- \frac{q^2}{4k_F^2} \right) \mbox{log} \left| \frac{1-q/2k_F}{1+q/2k_F} \right| \right]
\end{split}
\end{equation}
is the positive Lindhard function. 

The contribution from the ZS' diagram is obtained by switching indices 3 and 4 and by replacing $q$ by $q'$ in the loop integral. Explicitly,  
\begin{equation}
\begin{split}
    V^{\rm ind}_{ZS'}&= - U(q') \left[ (C_0^2+3\tilde{C}_0^2)\delta_{13} \delta_{24} - 6C_0 \tilde{C}_0 \delta_{14} \delta_{23})  \right]  \\
     &- U(q') \left[(4\tilde{C}_0^2 - 2C_0 \tilde{C}_0) \sigma_{14} \cdot \sigma_{23} + (2\tilde{C}_0^2 + 2C_0 \tilde{C}_0)\sigma_{13} \cdot \sigma_{24}) \right]\,.
\end{split}
\end{equation}

The calculation of the momentum-dependent part of the induced potential is similar but a bit more tedious and the analytic results involves a large number of terms. To obtain useful formula with fewer terms we present results for the spin singlet and spin-triplet contributions. These will require the second and fourth moments of the Lindhard function denoted $U_2$ and $U_4$.  $U_2$ is defined as follows:
\begin{equation}
\begin{split}
    U_2(q) &= -\frac{m}{2\pi^2q} \int_0^{k_F} \ell^3 d\ell \int_{-1}^1 \frac{d \cos{\theta}_{q\ell}}{\cos{\theta}_{q\ell} - q/2\ell} \\
    &= -\frac{mk_F^4}{2\pi^2 q} \left[ -\frac{q}{12k_F} - \frac{q^3}{16k_F^3} + \frac{1}{4} \left(1 - \frac{q^4}{16k_F^4} \right) \log \left| \frac{1-q/2k_F}{1+q/2k_F} \right| \right]
\end{split}
\end{equation}
$U_4$ is defined analagously and is given by:
\begin{equation}
    U_4(q) = -\frac{mk_F^6}{2\pi^2 q} \left[ -\frac{q}{30k_F} - \frac{q^3}{72k_F^3} -\frac{q^5}{96k_F^5} + \frac{1}{6} \left(1 - \frac{q^6}{64k_F^6} \right) \log \left| \frac{1-q/2k_F}{1+q/2k_F} \right| \right]
\end{equation}
Five momentum structures appear corresponding to the five pairings of the combinations of constants given above.  The momentum-dependent parts of $V_L$ and $V_R$ take the following form for the spin-independent terms.  The spin-dependent terms are analagous.
\begin{equation}
    \begin{split}
        V_L &\supset C_2 ((-\ell-k')^2+q^2) (\delta_{13} \delta_{ab} - \delta_{1b} \delta_{a3}) + C'_2 ((-\ell-k')^2-q^2) (\delta_{13} \delta_{ab} + \delta_{1b} \delta_{a3}) \\
        V_R &\supset C_2 ((\ell-k)^2+q^2) (\delta_{24} \delta_{ba} - \delta_{2a} \delta_{b4}) + C'_2 ((\ell-k)^2-q^2) (\delta_{24} \delta_{ba} + \delta_{2a} \delta_{b4})
    \end{split}
\end{equation}
The contributions of the momentum dependence to the induced potential are:
\begin{equation}
    \begin{split}
        \xi_a(q) &= \frac{1}{\beta} \sum_{\ell_0} \int \frac{d^3\ell}{(2\pi)^3} \Delta(\ell) \Delta(\ell+q) [2q^2 + (-\ell-k')^2+(\ell-k)^2] \\
        \xi_b(q) &= \frac{1}{\beta} \sum_{\ell_0} \int \frac{d^3\ell}{(2\pi)^3} \Delta(\ell) \Delta(\ell+q) [-2q^2 + (-\ell-k')^2+(\ell-k)^2] \\
        \xi_c(q) &= \frac{1}{\beta} \sum_{\ell_0} \int \frac{d^3\ell}{(2\pi)^3} \Delta(\ell) \Delta(\ell+q) [(-\ell-k')^2+q^2][(\ell-k)^2+q^2] \\
        \xi_d(q) &= \frac{1}{\beta} \sum_{\ell_0} \int \frac{d^3\ell}{(2\pi)^3} \Delta(\ell) \Delta(\ell+q) [(-\ell -k')^2-q^2][(\ell-k)^2-q^2]  \\
        \xi_e(q) &= \frac{1}{\beta} \sum_{\ell_0} \int \frac{d^3\ell}{(2\pi)^3} \Delta(\ell) \Delta(\ell+q) [((-\ell-k')^2-q^2)((\ell-k)^2+q^2)\\
        &+((\ell-k')^2+q^2)((\ell-k)^2-q^2)]
    \end{split}
\end{equation}
$\Delta(\ell) = (\ell_0 - \ell^2/2m)^{-1}$ is the fermion propagator. 
 After doing the loop integral, these give:
\begin{equation}
\begin{split}
    \xi_a (q) &= -\frac{2mk_F^3}{3\pi^2} - (q^2 + 2k_F^2) U(q) - 2U_2(q) \\
    \xi_b (q) &= -\frac{2mk_F^3}{3\pi^2} + (3q^2 - 2k_F^2) U(q) - 2U_2(q) \\ 
    \xi_c (q) &= -\frac{mk_F^3}{\pi^2} \left( \frac{11}{15}k_F^2 + \frac{7}{12}q^2 \right) -\left(k_F^4 + \frac{3}{2}q^2 k_F^2 + \frac{q^4}{8} \right) U(q) \\
    &-\frac{3}{2}q^2 U_2(q) - U_4(q) \\
    \xi_d (q) &= -\frac{mk_F^3}{\pi^2} \left( \frac{11}{15}k_F^2 - \frac{3}{4}q^2 \right) -\left(k_F^4 - \frac{5}{2}q^2 k_F^2 + \frac{17}{8}q^4 \right) U(q) \\
    &+\frac{5}{2}q^2 U_2(q) - U_4(q) \\
    \xi_e(q) &= -\frac{mk_F^3}{\pi^2} \left( \frac{22}{15}k_F^2 -\frac{q^2}{6} \right) - \left(2k_F^4-q^2 k_F^2 - \frac{7}{4} q^4 \right) + q^2 U_2(q) - 2 U_4(q)
\end{split}
\end{equation}
The total central induced potential in the spin triplet channel:
\begin{equation}
\begin{split}
    V^{\rm ind}_{S=1} &= -\bar{C}_0^2 (U(q)-U(q')) + \bar{C}_0 \bar{C}_2 (\xi_a(q)-\xi_a(q')) +\bar{C}_0 \bar{C}'_2 (\xi_b(q)-\xi_b(q')) \\
    &+ \bar{C}_2^2 (\xi_c(q)-\xi_c(q')) + 5 \bar{C}_2^{\prime 2} (\xi_d(q) - \xi_d(q')) +\bar{C}_2 \bar{C}'_2 (\xi_e(q)-\xi_e(q')) 
\end{split}
\end{equation}
The total central induced potential in the spin singlet channel:
\begin{equation}
\begin{split}
    V^{\rm ind}_{S=0} &= \bar{C}_0^2 \left( U(q)+U(q')\right) - \bar{C}_0 \bar{C}_2 (\xi_a(q)+\xi_a(q')) +3 \bar{C}_0 \bar{C}'_2 (\xi_b(q)+\xi_b(q')) \\
    &- \bar{C}_2^2 \left( \xi_c(q)+\xi_c(q') \right) + 3\left(\bar{C}_2^{\prime 2} (\xi_d(q) + \xi_d(q')) 
    +\bar{C}_2 \bar{C}'_2 (\xi_e(q)+\xi_e(q')\right)\,. 
\end{split}
\end{equation}
This gives s- and p-wave central potentials:
\begin{equation}
    \begin{split}
        ^1S_0 &: \bar{C}_0^2 \frac{mk_F}{3\pi^2} (1 + 2 \log 2) + mk_F^3 [\bar{C}_0 \bar{C}_2 \frac{2}{3\pi^2}(5+4\log{2}) \\
        &+ \bar{C}_0 \bar{C}'_2 \frac{2}{5\pi^2}(7-4\log{2})] +mk_F^5[\bar{C}_2^2 \frac{8}{315\pi^2}(277+96\log{2}) \\
        &- \bar{C}_2^{\prime 2} \frac{8}{105\pi^2} (43 + 24\log{2}) + \bar{C}_2 \bar{C}'_2 \frac{32}{105\pi^2}(37 + 6 \log{2})]
    \end{split}
\end{equation}
\begin{equation}
    \begin{split}
        ^3P_J &: \bar{C}_0^2 \frac{mk_F}{5\pi^2} (1 - 2 \log 2) + mk_F^3 [\bar{C}_0 \bar{C}_2 \frac{2}{105\pi^2}(59-68\log{2}) \\
        &- \bar{C}_0 \bar{C}'_2 \frac{2}{105\pi^2}(29+52\log{2})] +mk_F^5 [\bar{C}_2^2 \frac{16}{567\pi^2}(83-24\log{2}) \\
        &+\bar{C}_2^{\prime 2} \frac{64}{567\pi^2} (34 - 3\log{2}) - \bar{C}_2 \bar{C}'_2 \frac{16}{2835\pi^2}(523 + 204 \log{2})]
    \end{split}
\end{equation}

The spin-orbit potential gives an additional contribution to the p-waves:
\begin{equation}
    \begin{split}
         2\bar{C}'_2 V_{SO} \frac{1}{\beta} &\sum_{\ell_0} \int \frac{d^3\ell}{(2\pi)^3} \Delta(\ell) \Delta(\ell+q)  [((-\ell-k')^2-q^2) iq \times (\ell -k) \cdot (3\delta_{13} \sigma_{24} + \delta_{24} \sigma_{13}) \\
         &+((\ell-k)^2-q^2)iq \times (-\ell-k') \cdot (3\delta_{24} \sigma_{13} + \delta_{13} \sigma_{24})] \\
         &= \bar{C}'_2 V_{SO} i q \times q' \cdot (\sigma_{13} \delta_{24} + \sigma_{24} \delta_{13}) \xi_f(q)
    \end{split}
\end{equation}
The function $\xi_f(q)$ is given by:
\begin{equation}
    \xi_f(q) = -\frac{2mk_F^3}{3\pi^2} + (5q^2-4k_F^2) U(q)
\end{equation}
This gives a contribution to the p-waves after including the ZS' diagram:
\begin{equation}
    ^3P_J : [J(J+1)-4] \bar{C}'_2 V_{SO} \frac{32mk_F^5}{945\pi^2} (43 + 24 \log{2})
\end{equation}
We calculate the contribution of the tensor interaction only to $\mathcal{O}(mk_F^3)$. The term proportional to $C_0 V_T$ gives:
\begin{equation}
\begin{split}
    C_0 V_T \frac{1}{\beta} &\sum_{\ell_0} \int \frac{d^3\ell}{(2\pi)^3} \Delta(\ell) \Delta(\ell+q) [-\delta_{13} \delta_{24} ((-\ell-k')^2+(\ell-k)^2) - 2 q \cdot \sigma_{13} q \cdot \sigma_{24} \\
    &+ (-\ell-k') \cdot \sigma_{23} (-\ell-k') \cdot \sigma_{14} + (\ell-k) \cdot \sigma_{23} (\ell-k) \cdot \sigma_{14} ]
\end{split}
\end{equation}
Doing the calculation for the $\tilde{C}_0 V_T$ term gives $-3$ times the result for the term proportional to $C_0 V_T$ after reducing to spin singlet or triplet.  The potentials in these channels after including the ZS' diagram are given by:
\begin{equation}
    V^{\rm ind}_{S=0} = -\bar{C}_0 V_T (2q^2 U(q) + 2q^{\prime 2} U(q'))
\end{equation}
\begin{equation}
\begin{split}
    V^{\rm ind}_{S=1} &= \bar{C}_0 V_T \Bigl[ \frac{mk_F^3}{6\pi^2} (- \hat{q} \cdot \sigma_{13} \hat{q} \cdot \sigma_{24} + \hat{q}' \cdot \sigma_{13} \hat{q}' \cdot \sigma_{24}) + U(q) \left( \frac{3}{4} q \cdot \sigma_{13} q \cdot \sigma_{24} \right.  \\
    &- \left. \frac{1}{2} q' \cdot \sigma_{13} q' \cdot \sigma_{24} + \frac{q^2}{4} \right) - U(q') \left( \frac{3}{4} q' \cdot \sigma_{13} q' \cdot \sigma_{24} - \frac{1}{2} q \cdot \sigma_{13} q \cdot \sigma_{24} + \frac{q^{\prime 2}}{4} \right) \\
    &+ \bigl. U_2(q) (1 + \hat{q} \cdot \sigma_{13} \hat{q} \cdot \sigma_{24}) - U_2(q') (1 + \hat{q}' \cdot \sigma_{13} \hat{q}' \cdot \sigma_{24}) \Bigr]
\end{split}
\end{equation}
where we define the unit vector $\hat{q} = \vec{q}/|q|$. For the spin triplet, outgoing spin indices are exchanged on some terms to simplify the equations.  Doing the integrals gives:
\begin{equation}
    \begin{split}
        ^1S_0 &: - \bar{C}_0 V_T \frac{16mk_F^3}{15\pi^2} ( 2 + \log 2) \\
        ^3P_2 &: - \bar{C}_0 V_T \frac{4mk_F^3}{15\pi^2} (1 - \log 2) \\
        ^3P_1 &: - \bar{C}_0 V_T \frac{4mk_F^3}{21\pi^2} (4 + 5 \log 2) \\
        ^3P_0 &: \bar{C}_0 V_T \frac{2mk_F^3}{21\pi^2} (5 + 22 \log 2)
    \end{split}
\end{equation}

The part of the interaction proportional to $V_{SO}^2$ takes the form:
\begin{equation}
    \begin{split}
        -8V_{SO}^2 \frac{1}{\beta} &\sum_{\ell_0} \int \frac{d^3 \ell}{(2\pi)^3} \Delta(\ell) \Delta (\ell+q) [(q \times (-\ell -k') \cdot \sigma_{13})(q \times (\ell-k) \cdot \sigma_{24}) \\
        &+ \delta_{13} \delta_{24} (q \times (-\ell-k')) \cdot (q \times (\ell - k))] \\
        &= -\frac{2mk_F^3}{3\pi^2}(q^2(\sigma_{13} \cdot \sigma_{24} + 2\delta_{13} \delta_{24}) - (q \cdot \sigma_{13} )( q \cdot \sigma_{24})) + U(q) [q^4 \sigma_{13} \cdot \sigma_{24} \\
        &-q^2 (q \cdot \sigma_{13})(q \cdot \sigma_{24}) +2 (q \times q' \cdot \sigma_{13})(q \times q' \cdot \sigma_{24}) + 8\delta_{24} \delta_{24} q^2 k_F^2 ] \\
        &- U_2(q) [4 (q^2 \sigma_{13} \cdot \sigma_{24} - (q \cdot \sigma_{13} )(q \cdot \sigma_{24})) + 8  \delta_{13} \delta_{24} q^2]
    \end{split}
\end{equation}
A tedious calculation gives the following contribution:
\begin{equation}
    \begin{split}
        ^1S_0 &: V_{SO}^2 \frac{8mk_F^5}{35\pi^2}(17+16\log 2) \\
        ^3P_2 &: V_{SO}^2 \frac{128mk_F^5}{4725\pi^2}(43 + 24\log 2) \\
        ^3P_1 &: 0 \\
        ^3P_0 &: V_{SO}^2 \frac{64 mk_F^5}{945\pi^2} (43 + 24 \log 2)
    \end{split}
\end{equation}
\section{Induced interaction between quarks}
\label{App:QuarkInduced}
Treating quarks relativistically, the screening diagram is given by
\begin{equation}
    \begin{split}
        V^{\rm ind} &= g_V^2 (\bar{u}_3 \gamma_\mu u_1)  (\bar{u}_4 \gamma_\nu u_2) \left( \frac{E_k + m_s}{2m_s} \right)^2 \sum_{f,c} \frac{1}{\beta} \sum_{\ell_0} \int \frac{d^3 \ell}{(2\pi)^3} \\
        &\times \frac{\tr [\gamma^\mu (\slashed{\ell} + \slashed{q} + m_f) \gamma^\nu (\slashed{\ell} + m_f)]}{(\ell_0^2-\ell^2-m_f^2)(\ell_0^2-(\ell+q)^2-m_f^2)}
    \end{split}
\label{eq:v_ind_quark_0}
\end{equation}
where $f=u,d,s$ and $c=r,g,b$ denotes the flavor and color of quarks that appear in the particle-hole loop. Since the screening diagram is enhanced by the number of flavors and colors and the other diagrams are not, we calculate only this part of the potential.  We neglect anti-particle contributions by discarding the Matsubara sums that produce terms proportional to $(\exp [\beta (E+\mu)]+1)^{-1}$, which are negligible at small temperatures.  Doing the trace and noticing that $\bar{u}_3 \slashed{q} u_1 = \bar{u}_4 \slashed{q} u_2 = 0$, Eq.~\ref{eq:v_ind_quark_0} can be written as 
\begin{equation}
    V^{\rm ind} = g_V^2 (\bar{u}_3 \gamma_\mu u_1)  (\bar{u}_4 \gamma_\nu u_2) \left( \frac{E_k + m_s}{2m_s} \right)^2 \sum_{f,c} \int \frac{\ell  d\ell d\Omega_\ell}{4\pi^3 q E_\ell} \frac{\Theta(k_{fc} - \ell)}{\cos \theta_{q\ell} - q / 2\ell} (2 \ell^\mu \ell^\nu - g^{\mu \nu} \vec{\ell} \cdot \vec{q})
\end{equation}
$k_{fc}$ is the Fermi momentum of flavor $f$ and color $c$ with the normal subscript $F$ suppressed for readability. Since the Fermi momentum of strange quarks is approximately the same for all colors, also suppress the color label on $k_s$. Expanding to zeroth order in $k_s / m_s$, only the $\mu=\nu=0$ components contribute and we get:
\begin{equation}
    V^{\rm ind} = g_V^2 \delta_{13} \delta_{24} \frac{1}{2\pi^2 q} \sum_{f,c} \int \frac{\ell d\ell}{\sqrt{\ell^2 + m_f^2}} \frac{dc_{q\ell}}{\cos \theta_{q\ell}-q/2\ell} \Theta(k_{fc}-\ell) (2\ell^2 + 2 m_f^2 - 2\ell q \cos \theta_{q\ell})
\end{equation}
Setting $m_u=m_d=0$ and discarding components from the strange quark that are not proportional to $m_s$ gives:
\begin{equation}
    V^{\rm ind} = g_V^2 \delta_{13} \delta_{24} \sum_c \left[ \sum_{f=u,d} \left( -\frac{k_{fc}^2}{2\pi^2} +\frac{q^2}{2} U_0^{\rm rel} \left( \frac{q}{k_{fc}} \right) - 2k_{fc}^2 U_2^{\rm rel} \left( \frac{q}{k_{fc}} \right) \right) - 2 U(q) \right]
\end{equation}
where the mass in the Lindhard function $U(q)$ is the strange quark mass and we define relativistic dimensionless Lindhard functions in analogy with the relativistic ones (defining $\tilde{q} = q/k_{fc}$ to be distinguished from $\bar{q} = q/k_s$):
\begin{equation}
    \begin{split}
        U_0^{\rm rel} (\tilde{q}=q/k_{fc}) &= - \frac{1}{2\pi^2 \tilde{q}} \int d\bar{\ell} \log \left| \frac{1 - \tilde{q}/2\bar{\ell}}{1+\tilde{q}/2\bar{\ell}} \right|  \\
        &= \frac{1}{2\pi^2 \tilde{q}} \left[ \log \left| 1 + \frac{2}{\tilde{q}} \right| - \left(1 - \frac{\tilde{q}}{2} \right) \log \left| \frac{1-\tilde{q}/2}{1+\tilde{q}/2} \right| \right]\\
        U_2^{\rm rel} (\tilde{q}) &= - \frac{1}{2\pi^2 \tilde{q}} \int \bar{\ell}^2 d\bar{\ell} \log \left| \frac{1 - \tilde{q}/2\bar{\ell}}{1+\tilde{q}/2\bar{\ell}} \right|  \\ 
        &= \frac{1}{6\pi^2 \tilde{q}} \left[ \frac{\tilde{q}}{2} + \frac{\tilde{q}^3}{4} \log \left|1 + \frac{2}{\tilde{q}} \right| - \left(1 - \frac{\tilde{q}^3}{8} \right) \log \left| \frac{1-\tilde{q}/2}{1+\tilde{q}/2} \right| \right]
    \end{split}
\end{equation}
Analytical expressions for s- and p-wave potentials can easily be found with Mathematica or equivalent, but are long and unenlightening so we do not reproduce them here.

\bibliographystyle{apsrev4-2}
\bibliography{kl_refs}

\begin{thebibliography}{50}%
\makeatletter
\providecommand \@ifxundefined [1]{%
 \@ifx{#1\undefined}
}%
\providecommand \@ifnum [1]{%
 \ifnum #1\expandafter \@firstoftwo
 \else \expandafter \@secondoftwo
 \fi
}%
\providecommand \@ifx [1]{%
 \ifx #1\expandafter \@firstoftwo
 \else \expandafter \@secondoftwo
 \fi
}%
\providecommand \natexlab [1]{#1}%
\providecommand \enquote  [1]{``#1''}%
\providecommand \bibnamefont  [1]{#1}%
\providecommand \bibfnamefont [1]{#1}%
\providecommand \citenamefont [1]{#1}%
\providecommand \href@noop [0]{\@secondoftwo}%
\providecommand \href [0]{\begingroup \@sanitize@url \@href}%
\providecommand \@href[1]{\@@startlink{#1}\@@href}%
\providecommand \@@href[1]{\endgroup#1\@@endlink}%
\providecommand \@sanitize@url [0]{\catcode `\\12\catcode `\$12\catcode `\&12\catcode `\#12\catcode `\^12\catcode `\_12\catcode `\%12\relax}%
\providecommand \@@startlink[1]{}%
\providecommand \@@endlink[0]{}%
\providecommand \url  [0]{\begingroup\@sanitize@url \@url }%
\providecommand \@url [1]{\endgroup\@href {#1}{\urlprefix }}%
\providecommand \urlprefix  [0]{URL }%
\providecommand \Eprint [0]{\href }%
\providecommand \doibase [0]{https://doi.org/}%
\providecommand \selectlanguage [0]{\@gobble}%
\providecommand \bibinfo  [0]{\@secondoftwo}%
\providecommand \bibfield  [0]{\@secondoftwo}%
\providecommand \translation [1]{[#1]}%
\providecommand \BibitemOpen [0]{}%
\providecommand \bibitemStop [0]{}%
\providecommand \bibitemNoStop [0]{.\EOS\space}%
\providecommand \EOS [0]{\spacefactor3000\relax}%
\providecommand \BibitemShut  [1]{\csname bibitem#1\endcsname}%
\let\auto@bib@innerbib\@empty
\bibitem [{\citenamefont {Cromartie}\ \emph {et~al.}(2019)\citenamefont {Cromartie} \emph {et~al.}}]{NANOGrav:2019jur}%
  \BibitemOpen
  \bibfield  {author} {\bibinfo {author} {\bibfnamefont {H.~T.}\ \bibnamefont {Cromartie}} \emph {et~al.} (\bibinfo {collaboration} {NANOGrav}),\ }\href {https://doi.org/10.1038/s41550-019-0880-2} {\bibfield  {journal} {\bibinfo  {journal} {Nature Astron.}\ }\textbf {\bibinfo {volume} {4}},\ \bibinfo {pages} {72} (\bibinfo {year} {2019})},\ \Eprint {https://arxiv.org/abs/1904.06759} {arXiv:1904.06759 [astro-ph.HE]} \BibitemShut {NoStop}%
\bibitem [{\citenamefont {Antoniadis}\ \emph {et~al.}(2013)\citenamefont {Antoniadis} \emph {et~al.}}]{Antoniadis:2013pzd}%
  \BibitemOpen
  \bibfield  {author} {\bibinfo {author} {\bibfnamefont {J.}~\bibnamefont {Antoniadis}} \emph {et~al.},\ }\href {https://doi.org/10.1126/science.1233232} {\bibfield  {journal} {\bibinfo  {journal} {Science}\ }\textbf {\bibinfo {volume} {340}},\ \bibinfo {pages} {6131} (\bibinfo {year} {2013})},\ \Eprint {https://arxiv.org/abs/1304.6875} {arXiv:1304.6875 [astro-ph.HE]} \BibitemShut {NoStop}%
\bibitem [{\citenamefont {Demorest}\ \emph {et~al.}(2010)\citenamefont {Demorest}, \citenamefont {Pennucci}, \citenamefont {Ransom}, \citenamefont {Roberts},\ and\ \citenamefont {Hessels}}]{Demorest:2010bx}%
  \BibitemOpen
  \bibfield  {author} {\bibinfo {author} {\bibfnamefont {P.}~\bibnamefont {Demorest}}, \bibinfo {author} {\bibfnamefont {T.}~\bibnamefont {Pennucci}}, \bibinfo {author} {\bibfnamefont {S.}~\bibnamefont {Ransom}}, \bibinfo {author} {\bibfnamefont {M.}~\bibnamefont {Roberts}},\ and\ \bibinfo {author} {\bibfnamefont {J.}~\bibnamefont {Hessels}},\ }\href {https://doi.org/10.1038/nature09466} {\bibfield  {journal} {\bibinfo  {journal} {Nature}\ }\textbf {\bibinfo {volume} {467}},\ \bibinfo {pages} {1081} (\bibinfo {year} {2010})},\ \Eprint {https://arxiv.org/abs/1010.5788} {arXiv:1010.5788 [astro-ph.HE]} \BibitemShut {NoStop}%
\bibitem [{\citenamefont {Abbott}\ \emph {et~al.}(2018)\citenamefont {Abbott} \emph {et~al.}}]{LIGOScientific:2018cki}%
  \BibitemOpen
  \bibfield  {author} {\bibinfo {author} {\bibfnamefont {B.~P.}\ \bibnamefont {Abbott}} \emph {et~al.} (\bibinfo {collaboration} {LIGO Scientific, Virgo}),\ }\href {https://doi.org/10.1103/PhysRevLett.121.161101} {\bibfield  {journal} {\bibinfo  {journal} {Phys. Rev. Lett.}\ }\textbf {\bibinfo {volume} {121}},\ \bibinfo {pages} {161101} (\bibinfo {year} {2018})},\ \Eprint {https://arxiv.org/abs/1805.11581} {arXiv:1805.11581 [gr-qc]} \BibitemShut {NoStop}%
\bibitem [{\citenamefont {De}\ \emph {et~al.}(2018)\citenamefont {De}, \citenamefont {Finstad}, \citenamefont {Lattimer}, \citenamefont {Brown}, \citenamefont {Berger},\ and\ \citenamefont {Biwer}}]{De:2018uhw}%
  \BibitemOpen
  \bibfield  {author} {\bibinfo {author} {\bibfnamefont {S.}~\bibnamefont {De}}, \bibinfo {author} {\bibfnamefont {D.}~\bibnamefont {Finstad}}, \bibinfo {author} {\bibfnamefont {J.~M.}\ \bibnamefont {Lattimer}}, \bibinfo {author} {\bibfnamefont {D.~A.}\ \bibnamefont {Brown}}, \bibinfo {author} {\bibfnamefont {E.}~\bibnamefont {Berger}},\ and\ \bibinfo {author} {\bibfnamefont {C.~M.}\ \bibnamefont {Biwer}},\ }\href {https://doi.org/10.1103/PhysRevLett.121.091102} {\bibfield  {journal} {\bibinfo  {journal} {Phys. Rev. Lett.}\ }\textbf {\bibinfo {volume} {121}},\ \bibinfo {pages} {091102} (\bibinfo {year} {2018})},\ \bibinfo {note} {[Erratum: Phys.Rev.Lett. 121, 259902 (2018)]},\ \Eprint {https://arxiv.org/abs/1804.08583} {arXiv:1804.08583 [astro-ph.HE]} \BibitemShut {NoStop}%
\bibitem [{\citenamefont {Capano}\ \emph {et~al.}(2020)\citenamefont {Capano}, \citenamefont {Tews}, \citenamefont {Brown}, \citenamefont {Margalit}, \citenamefont {De}, \citenamefont {Kumar}, \citenamefont {Brown}, \citenamefont {Krishnan},\ and\ \citenamefont {Reddy}}]{Capano:2019eae}%
  \BibitemOpen
  \bibfield  {author} {\bibinfo {author} {\bibfnamefont {C.~D.}\ \bibnamefont {Capano}}, \bibinfo {author} {\bibfnamefont {I.}~\bibnamefont {Tews}}, \bibinfo {author} {\bibfnamefont {S.~M.}\ \bibnamefont {Brown}}, \bibinfo {author} {\bibfnamefont {B.}~\bibnamefont {Margalit}}, \bibinfo {author} {\bibfnamefont {S.}~\bibnamefont {De}}, \bibinfo {author} {\bibfnamefont {S.}~\bibnamefont {Kumar}}, \bibinfo {author} {\bibfnamefont {D.~A.}\ \bibnamefont {Brown}}, \bibinfo {author} {\bibfnamefont {B.}~\bibnamefont {Krishnan}},\ and\ \bibinfo {author} {\bibfnamefont {S.}~\bibnamefont {Reddy}},\ }\href {https://doi.org/10.1038/s41550-020-1014-6} {\bibfield  {journal} {\bibinfo  {journal} {Nature Astron.}\ }\textbf {\bibinfo {volume} {4}},\ \bibinfo {pages} {625} (\bibinfo {year} {2020})},\ \Eprint {https://arxiv.org/abs/1908.10352} {arXiv:1908.10352 [astro-ph.HE]} \BibitemShut {NoStop}%
\bibitem [{\citenamefont {Miller}\ \emph {et~al.}(2019)\citenamefont {Miller} \emph {et~al.}}]{Miller:2019cac}%
  \BibitemOpen
  \bibfield  {author} {\bibinfo {author} {\bibfnamefont {M.~C.}\ \bibnamefont {Miller}} \emph {et~al.},\ }\href {https://doi.org/10.3847/2041-8213/ab50c5} {\bibfield  {journal} {\bibinfo  {journal} {Astrophys. J. Lett.}\ }\textbf {\bibinfo {volume} {887}},\ \bibinfo {pages} {L24} (\bibinfo {year} {2019})},\ \Eprint {https://arxiv.org/abs/1912.05705} {arXiv:1912.05705 [astro-ph.HE]} \BibitemShut {NoStop}%
\bibitem [{\citenamefont {Riley}\ \emph {et~al.}(2019)\citenamefont {Riley} \emph {et~al.}}]{Riley:2019yda}%
  \BibitemOpen
  \bibfield  {author} {\bibinfo {author} {\bibfnamefont {T.~E.}\ \bibnamefont {Riley}} \emph {et~al.},\ }\href {https://doi.org/10.3847/2041-8213/ab481c} {\bibfield  {journal} {\bibinfo  {journal} {Astrophys. J. Lett.}\ }\textbf {\bibinfo {volume} {887}},\ \bibinfo {pages} {L21} (\bibinfo {year} {2019})},\ \Eprint {https://arxiv.org/abs/1912.05702} {arXiv:1912.05702 [astro-ph.HE]} \BibitemShut {NoStop}%
\bibitem [{\citenamefont {Hebeler}\ and\ \citenamefont {Schwenk}(2010)}]{Hebeler:2009iv}%
  \BibitemOpen
  \bibfield  {author} {\bibinfo {author} {\bibfnamefont {K.}~\bibnamefont {Hebeler}}\ and\ \bibinfo {author} {\bibfnamefont {A.}~\bibnamefont {Schwenk}},\ }\href {https://doi.org/10.1103/PhysRevC.82.014314} {\bibfield  {journal} {\bibinfo  {journal} {Phys. Rev. C}\ }\textbf {\bibinfo {volume} {82}},\ \bibinfo {pages} {014314} (\bibinfo {year} {2010})},\ \Eprint {https://arxiv.org/abs/0911.0483} {arXiv:0911.0483 [nucl-th]} \BibitemShut {NoStop}%
\bibitem [{\citenamefont {Gandolfi}\ \emph {et~al.}(2014)\citenamefont {Gandolfi}, \citenamefont {Carlson}, \citenamefont {Reddy}, \citenamefont {Steiner},\ and\ \citenamefont {Wiringa}}]{Gandolfi:2013baa}%
  \BibitemOpen
  \bibfield  {author} {\bibinfo {author} {\bibfnamefont {S.}~\bibnamefont {Gandolfi}}, \bibinfo {author} {\bibfnamefont {J.}~\bibnamefont {Carlson}}, \bibinfo {author} {\bibfnamefont {S.}~\bibnamefont {Reddy}}, \bibinfo {author} {\bibfnamefont {A.~W.}\ \bibnamefont {Steiner}},\ and\ \bibinfo {author} {\bibfnamefont {R.~B.}\ \bibnamefont {Wiringa}},\ }\href {https://doi.org/10.1140/epja/i2014-14010-5} {\bibfield  {journal} {\bibinfo  {journal} {Eur. Phys. J. A}\ }\textbf {\bibinfo {volume} {50}},\ \bibinfo {pages} {10} (\bibinfo {year} {2014})},\ \Eprint {https://arxiv.org/abs/1307.5815} {arXiv:1307.5815 [nucl-th]} \BibitemShut {NoStop}%
\bibitem [{\citenamefont {Tews}\ \emph {et~al.}(2013)\citenamefont {Tews}, \citenamefont {Kr\"uger}, \citenamefont {Hebeler},\ and\ \citenamefont {Schwenk}}]{Tews:2012fj}%
  \BibitemOpen
  \bibfield  {author} {\bibinfo {author} {\bibfnamefont {I.}~\bibnamefont {Tews}}, \bibinfo {author} {\bibfnamefont {T.}~\bibnamefont {Kr\"uger}}, \bibinfo {author} {\bibfnamefont {K.}~\bibnamefont {Hebeler}},\ and\ \bibinfo {author} {\bibfnamefont {A.}~\bibnamefont {Schwenk}},\ }\href {https://doi.org/10.1103/PhysRevLett.110.032504} {\bibfield  {journal} {\bibinfo  {journal} {Phys. Rev. Lett.}\ }\textbf {\bibinfo {volume} {110}},\ \bibinfo {pages} {032504} (\bibinfo {year} {2013})},\ \Eprint {https://arxiv.org/abs/1206.0025} {arXiv:1206.0025 [nucl-th]} \BibitemShut {NoStop}%
\bibitem [{\citenamefont {Lynn}\ \emph {et~al.}(2016)\citenamefont {Lynn}, \citenamefont {Tews}, \citenamefont {Carlson}, \citenamefont {Gandolfi}, \citenamefont {Gezerlis}, \citenamefont {Schmidt},\ and\ \citenamefont {Schwenk}}]{Lynn:2015jua}%
  \BibitemOpen
  \bibfield  {author} {\bibinfo {author} {\bibfnamefont {J.~E.}\ \bibnamefont {Lynn}}, \bibinfo {author} {\bibfnamefont {I.}~\bibnamefont {Tews}}, \bibinfo {author} {\bibfnamefont {J.}~\bibnamefont {Carlson}}, \bibinfo {author} {\bibfnamefont {S.}~\bibnamefont {Gandolfi}}, \bibinfo {author} {\bibfnamefont {A.}~\bibnamefont {Gezerlis}}, \bibinfo {author} {\bibfnamefont {K.~E.}\ \bibnamefont {Schmidt}},\ and\ \bibinfo {author} {\bibfnamefont {A.}~\bibnamefont {Schwenk}},\ }\href {https://doi.org/10.1103/PhysRevLett.116.062501} {\bibfield  {journal} {\bibinfo  {journal} {Phys. Rev. Lett.}\ }\textbf {\bibinfo {volume} {116}},\ \bibinfo {pages} {062501} (\bibinfo {year} {2016})},\ \Eprint {https://arxiv.org/abs/1509.03470} {arXiv:1509.03470 [nucl-th]} \BibitemShut {NoStop}%
\bibitem [{\citenamefont {Drischler}\ \emph {et~al.}(2020)\citenamefont {Drischler}, \citenamefont {Furnstahl}, \citenamefont {Melendez},\ and\ \citenamefont {Phillips}}]{Drischler:2020hwi}%
  \BibitemOpen
  \bibfield  {author} {\bibinfo {author} {\bibfnamefont {C.}~\bibnamefont {Drischler}}, \bibinfo {author} {\bibfnamefont {R.~J.}\ \bibnamefont {Furnstahl}}, \bibinfo {author} {\bibfnamefont {J.~A.}\ \bibnamefont {Melendez}},\ and\ \bibinfo {author} {\bibfnamefont {D.~R.}\ \bibnamefont {Phillips}},\ }\href {https://doi.org/10.1103/PhysRevLett.125.202702} {\bibfield  {journal} {\bibinfo  {journal} {Phys. Rev. Lett.}\ }\textbf {\bibinfo {volume} {125}},\ \bibinfo {pages} {202702} (\bibinfo {year} {2020})},\ \Eprint {https://arxiv.org/abs/2004.07232} {arXiv:2004.07232 [nucl-th]} \BibitemShut {NoStop}%
\bibitem [{\citenamefont {Tews}\ \emph {et~al.}(2018)\citenamefont {Tews}, \citenamefont {Carlson}, \citenamefont {Gandolfi},\ and\ \citenamefont {Reddy}}]{Tews:2018kmu}%
  \BibitemOpen
  \bibfield  {author} {\bibinfo {author} {\bibfnamefont {I.}~\bibnamefont {Tews}}, \bibinfo {author} {\bibfnamefont {J.}~\bibnamefont {Carlson}}, \bibinfo {author} {\bibfnamefont {S.}~\bibnamefont {Gandolfi}},\ and\ \bibinfo {author} {\bibfnamefont {S.}~\bibnamefont {Reddy}},\ }\href {https://doi.org/10.3847/1538-4357/aac267} {\bibfield  {journal} {\bibinfo  {journal} {Astrophys. J.}\ }\textbf {\bibinfo {volume} {860}},\ \bibinfo {pages} {149} (\bibinfo {year} {2018})},\ \Eprint {https://arxiv.org/abs/1801.01923} {arXiv:1801.01923 [nucl-th]} \BibitemShut {NoStop}%
\bibitem [{\citenamefont {Kohn}\ and\ \citenamefont {Luttinger}(1965)}]{KohnLuttinger1965PRL}%
  \BibitemOpen
  \bibfield  {author} {\bibinfo {author} {\bibfnamefont {W.}~\bibnamefont {Kohn}}\ and\ \bibinfo {author} {\bibfnamefont {J.~M.}\ \bibnamefont {Luttinger}},\ }\href {https://doi.org/10.1103/PhysRevLett.15.524} {\bibfield  {journal} {\bibinfo  {journal} {Phys. Rev. Lett.}\ }\textbf {\bibinfo {volume} {15}},\ \bibinfo {pages} {524} (\bibinfo {year} {1965})}\BibitemShut {NoStop}%
\bibitem [{\citenamefont {Kagan}(2013)}]{Kagan:2013}%
  \BibitemOpen
  \bibfield  {author} {\bibinfo {author} {\bibfnamefont {M.~Y.}\ \bibnamefont {Kagan}},\ }\href {https://doi.org/10.1007/978-94-007-6961-8} {\emph {\bibinfo {title} {Modern trends in Superconductivity and Superfluidity}}}\ (\bibinfo  {publisher} {Springer Netherlands},\ \bibinfo {year} {2013})\BibitemShut {NoStop}%
\bibitem [{\citenamefont {Fay}\ and\ \citenamefont {Layzer}(1968)}]{Fay:1968}%
  \BibitemOpen
  \bibfield  {author} {\bibinfo {author} {\bibfnamefont {D.}~\bibnamefont {Fay}}\ and\ \bibinfo {author} {\bibfnamefont {A.}~\bibnamefont {Layzer}},\ }\href {https://doi.org/10.1103/PhysRevLett.20.187} {\bibfield  {journal} {\bibinfo  {journal} {Phys. Rev. Lett.}\ }\textbf {\bibinfo {volume} {20}},\ \bibinfo {pages} {187} (\bibinfo {year} {1968})}\BibitemShut {NoStop}%
\bibitem [{\citenamefont {Pines}(1971)}]{Pines:1971}%
  \BibitemOpen
  \bibfield  {author} {\bibinfo {author} {\bibfnamefont {D.}~\bibnamefont {Pines}},\ }in\ \href@noop {} {\emph {\bibinfo {booktitle} {Proc. XIth Intern. Conf. on Low temperature physics}}}\ (\bibinfo  {publisher} {Academic Press of Japan, Tokyo},\ \bibinfo {year} {1971})\ p.~\bibinfo {pages} {10}\BibitemShut {NoStop}%
\bibitem [{\citenamefont {Clark}\ \emph {et~al.}(1976)\citenamefont {Clark}, \citenamefont {Källman}, \citenamefont {Yang},\ and\ \citenamefont {Chakkalakal}}]{Clark:1976}%
  \BibitemOpen
  \bibfield  {author} {\bibinfo {author} {\bibfnamefont {J.}~\bibnamefont {Clark}}, \bibinfo {author} {\bibfnamefont {C.-G.}\ \bibnamefont {Källman}}, \bibinfo {author} {\bibfnamefont {C.-H.}\ \bibnamefont {Yang}},\ and\ \bibinfo {author} {\bibfnamefont {D.}~\bibnamefont {Chakkalakal}},\ }\href {https://doi.org/https://doi.org/10.1016/0370-2693(76)90580-3} {\bibfield  {journal} {\bibinfo  {journal} {Physics Letters B}\ }\textbf {\bibinfo {volume} {61}},\ \bibinfo {pages} {331} (\bibinfo {year} {1976})}\BibitemShut {NoStop}%
\bibitem [{\citenamefont {Dean}\ and\ \citenamefont {Hjorth-Jensen}(2003)}]{Dean:2002zx}%
  \BibitemOpen
  \bibfield  {author} {\bibinfo {author} {\bibfnamefont {D.~J.}\ \bibnamefont {Dean}}\ and\ \bibinfo {author} {\bibfnamefont {M.}~\bibnamefont {Hjorth-Jensen}},\ }\href {https://doi.org/10.1103/RevModPhys.75.607} {\bibfield  {journal} {\bibinfo  {journal} {Rev. Mod. Phys.}\ }\textbf {\bibinfo {volume} {75}},\ \bibinfo {pages} {607} (\bibinfo {year} {2003})},\ \Eprint {https://arxiv.org/abs/nucl-th/0210033} {arXiv:nucl-th/0210033} \BibitemShut {NoStop}%
\bibitem [{\citenamefont {Gezerlis}\ \emph {et~al.}(2014)\citenamefont {Gezerlis}, \citenamefont {Pethick},\ and\ \citenamefont {Schwenk}}]{Gezerlis:2014efa}%
  \BibitemOpen
  \bibfield  {author} {\bibinfo {author} {\bibfnamefont {A.}~\bibnamefont {Gezerlis}}, \bibinfo {author} {\bibfnamefont {C.~J.}\ \bibnamefont {Pethick}},\ and\ \bibinfo {author} {\bibfnamefont {A.}~\bibnamefont {Schwenk}},\ }\href@noop {} {\  (\bibinfo {year} {2014})},\ \Eprint {https://arxiv.org/abs/1406.6109} {arXiv:1406.6109 [nucl-th]} \BibitemShut {NoStop}%
\bibitem [{\citenamefont {Gorkov}\ and\ \citenamefont {Melik-Barkhudarov}(1961)}]{Gorkov:1961}%
  \BibitemOpen
  \bibfield  {author} {\bibinfo {author} {\bibfnamefont {L.~P.}\ \bibnamefont {Gorkov}}\ and\ \bibinfo {author} {\bibfnamefont {T.~K.}\ \bibnamefont {Melik-Barkhudarov}},\ }\href@noop {} {\bibfield  {journal} {\bibinfo  {journal} {Sov. Phys. JETP}\ }\textbf {\bibinfo {volume} {13}},\ \bibinfo {pages} {1018} (\bibinfo {year} {1961})}\BibitemShut {NoStop}%
\bibitem [{\citenamefont {Schwenk}\ and\ \citenamefont {Friman}(2004)}]{Schwenk:2003bc}%
  \BibitemOpen
  \bibfield  {author} {\bibinfo {author} {\bibfnamefont {A.}~\bibnamefont {Schwenk}}\ and\ \bibinfo {author} {\bibfnamefont {B.}~\bibnamefont {Friman}},\ }\href {https://doi.org/10.1103/PhysRevLett.92.082501} {\bibfield  {journal} {\bibinfo  {journal} {Phys. Rev. Lett.}\ }\textbf {\bibinfo {volume} {92}},\ \bibinfo {pages} {082501} (\bibinfo {year} {2004})},\ \Eprint {https://arxiv.org/abs/nucl-th/0307089} {arXiv:nucl-th/0307089} \BibitemShut {NoStop}%
\bibitem [{\citenamefont {Baranov}\ \emph {et~al.}(1992)\citenamefont {Baranov}, \citenamefont {Chubukov},\ and\ \citenamefont {Kagan}}]{BaranovKL}%
  \BibitemOpen
  \bibfield  {author} {\bibinfo {author} {\bibfnamefont {M.~A.}\ \bibnamefont {Baranov}}, \bibinfo {author} {\bibfnamefont {A.~V.}\ \bibnamefont {Chubukov}},\ and\ \bibinfo {author} {\bibfnamefont {M.~Y.}\ \bibnamefont {Kagan}},\ }\href@noop {} {\bibfield  {journal} {\bibinfo  {journal} {Int. J. Mod. Phys. B}\ }\textbf {\bibinfo {volume} {6}},\ \bibinfo {pages} {2471} (\bibinfo {year} {1992})}\BibitemShut {NoStop}%
\bibitem [{\citenamefont {Gonz\'alez}(2008)}]{Gonzalez2008PRB}%
  \BibitemOpen
  \bibfield  {author} {\bibinfo {author} {\bibfnamefont {J.}~\bibnamefont {Gonz\'alez}},\ }\href {https://doi.org/10.1103/PhysRevB.78.205431} {\bibfield  {journal} {\bibinfo  {journal} {Phys. Rev. B}\ }\textbf {\bibinfo {volume} {78}},\ \bibinfo {pages} {205431} (\bibinfo {year} {2008})}\BibitemShut {NoStop}%
\bibitem [{\citenamefont {Nandkishore}\ \emph {et~al.}(2014)\citenamefont {Nandkishore}, \citenamefont {Thomale},\ and\ \citenamefont {Chubukov}}]{Nandkishoreetal2014PRB}%
  \BibitemOpen
  \bibfield  {author} {\bibinfo {author} {\bibfnamefont {R.}~\bibnamefont {Nandkishore}}, \bibinfo {author} {\bibfnamefont {R.}~\bibnamefont {Thomale}},\ and\ \bibinfo {author} {\bibfnamefont {A.~V.}\ \bibnamefont {Chubukov}},\ }\href {https://doi.org/10.1103/PhysRevB.89.144501} {\bibfield  {journal} {\bibinfo  {journal} {Phys. Rev. B}\ }\textbf {\bibinfo {volume} {89}},\ \bibinfo {pages} {144501} (\bibinfo {year} {2014})}\BibitemShut {NoStop}%
\bibitem [{\citenamefont {Gonz\'alez}\ and\ \citenamefont {Stauber}(2019)}]{Gonzalez2019PRL}%
  \BibitemOpen
  \bibfield  {author} {\bibinfo {author} {\bibfnamefont {J.}~\bibnamefont {Gonz\'alez}}\ and\ \bibinfo {author} {\bibfnamefont {T.}~\bibnamefont {Stauber}},\ }\href {https://doi.org/10.1103/PhysRevLett.122.026801} {\bibfield  {journal} {\bibinfo  {journal} {Phys. Rev. Lett.}\ }\textbf {\bibinfo {volume} {122}},\ \bibinfo {pages} {026801} (\bibinfo {year} {2019})}\BibitemShut {NoStop}%
\bibitem [{\citenamefont {Efremov}\ \emph {et~al.}(2000)\citenamefont {Efremov}, \citenamefont {Mar'enko}, \citenamefont {Baranov},\ and\ \citenamefont {Kagan}}]{Efremov_2000}%
  \BibitemOpen
  \bibfield  {author} {\bibinfo {author} {\bibfnamefont {D.~V.}\ \bibnamefont {Efremov}}, \bibinfo {author} {\bibfnamefont {M.~S.}\ \bibnamefont {Mar'enko}}, \bibinfo {author} {\bibfnamefont {M.~A.}\ \bibnamefont {Baranov}},\ and\ \bibinfo {author} {\bibfnamefont {M.~Y.}\ \bibnamefont {Kagan}},\ }\href {https://doi.org/10.1134/1.559173} {\bibfield  {journal} {\bibinfo  {journal} {J. Exp. Theor. Phys.}\ }\textbf {\bibinfo {volume} {90}},\ \bibinfo {pages} {861} (\bibinfo {year} {2000})},\ \Eprint {https://arxiv.org/abs/arXiv:cond-mat/0007334} {arXiv:cond-mat/0007334} \BibitemShut {NoStop}%
\bibitem [{\citenamefont {Maiti}\ and\ \citenamefont {Chubukov}(2013)}]{MaitiChubukov2013AIPKLReview}%
  \BibitemOpen
  \bibfield  {author} {\bibinfo {author} {\bibfnamefont {S.}~\bibnamefont {Maiti}}\ and\ \bibinfo {author} {\bibfnamefont {A.~V.}\ \bibnamefont {Chubukov}},\ }\href@noop {} {\bibfield  {journal} {\bibinfo  {journal} {AIP Conference Proceedings}\ }\textbf {\bibinfo {volume} {1550}},\ \bibinfo {pages} {3} (\bibinfo {year} {2013})}\BibitemShut {NoStop}%
\bibitem [{\citenamefont {Kagan}\ and\ \citenamefont {Chubukov}(1988)}]{Kagan:1988}%
  \BibitemOpen
  \bibfield  {author} {\bibinfo {author} {\bibfnamefont {M.~Y.}\ \bibnamefont {Kagan}}\ and\ \bibinfo {author} {\bibfnamefont {A.}~\bibnamefont {Chubukov}},\ }\href@noop {} {\bibfield  {journal} {\bibinfo  {journal} {JETP Lett}\ }\textbf {\bibinfo {volume} {47}},\ \bibinfo {pages} {614} (\bibinfo {year} {1988})}\BibitemShut {NoStop}%
\bibitem [{\citenamefont {Friman}\ and\ \citenamefont {Weise}(2019)}]{Friman:2019ncm}%
  \BibitemOpen
  \bibfield  {author} {\bibinfo {author} {\bibfnamefont {B.}~\bibnamefont {Friman}}\ and\ \bibinfo {author} {\bibfnamefont {W.}~\bibnamefont {Weise}},\ }\href {https://doi.org/10.1103/PhysRevC.100.065807} {\bibfield  {journal} {\bibinfo  {journal} {Phys. Rev. C}\ }\textbf {\bibinfo {volume} {100}},\ \bibinfo {pages} {065807} (\bibinfo {year} {2019})},\ \Eprint {https://arxiv.org/abs/1908.09722} {arXiv:1908.09722 [nucl-th]} \BibitemShut {NoStop}%
\bibitem [{\citenamefont {Baldo}\ \emph {et~al.}(1998)\citenamefont {Baldo}, \citenamefont {Elgar\o{}y}, \citenamefont {Engvik}, \citenamefont {Hjorth-Jensen},\ and\ \citenamefont {Schulze}}]{Baldoetal1998PRCNeutronGaps}%
  \BibitemOpen
  \bibfield  {author} {\bibinfo {author} {\bibfnamefont {M.}~\bibnamefont {Baldo}}, \bibinfo {author} {\bibfnamefont {O.}~\bibnamefont {Elgar\o{}y}}, \bibinfo {author} {\bibfnamefont {L.}~\bibnamefont {Engvik}}, \bibinfo {author} {\bibfnamefont {M.}~\bibnamefont {Hjorth-Jensen}},\ and\ \bibinfo {author} {\bibfnamefont {H.-J.}\ \bibnamefont {Schulze}},\ }\href {https://doi.org/10.1103/PhysRevC.58.1921} {\bibfield  {journal} {\bibinfo  {journal} {Phys. Rev. C}\ }\textbf {\bibinfo {volume} {58}},\ \bibinfo {pages} {1921} (\bibinfo {year} {1998})}\BibitemShut {NoStop}%
\bibitem [{\citenamefont {Drischler}\ \emph {et~al.}(2017)\citenamefont {Drischler}, \citenamefont {Kr\"uger}, \citenamefont {Hebeler},\ and\ \citenamefont {Schwenk}}]{Drischler:2016cpy}%
  \BibitemOpen
  \bibfield  {author} {\bibinfo {author} {\bibfnamefont {C.}~\bibnamefont {Drischler}}, \bibinfo {author} {\bibfnamefont {T.}~\bibnamefont {Kr\"uger}}, \bibinfo {author} {\bibfnamefont {K.}~\bibnamefont {Hebeler}},\ and\ \bibinfo {author} {\bibfnamefont {A.}~\bibnamefont {Schwenk}},\ }\href {https://doi.org/10.1103/PhysRevC.95.024302} {\bibfield  {journal} {\bibinfo  {journal} {Phys. Rev. C}\ }\textbf {\bibinfo {volume} {95}},\ \bibinfo {pages} {024302} (\bibinfo {year} {2017})},\ \Eprint {https://arxiv.org/abs/1610.05213} {arXiv:1610.05213 [nucl-th]} \BibitemShut {NoStop}%
\bibitem [{\citenamefont {Pethick}\ and\ \citenamefont {Ravenhall}(1991)}]{Pethick:1991}%
  \BibitemOpen
  \bibfield  {author} {\bibinfo {author} {\bibfnamefont {C.~J.}\ \bibnamefont {Pethick}}\ and\ \bibinfo {author} {\bibfnamefont {D.~G.}\ \bibnamefont {Ravenhall}},\ }\href {https://doi.org/https://doi.org/10.1111/j.1749-6632.1991.tb32200.x} {\bibfield  {journal} {\bibinfo  {journal} {Annals of the New York Academy of Sciences}\ }\textbf {\bibinfo {volume} {647}},\ \bibinfo {pages} {503} (\bibinfo {year} {1991})},\ \Eprint {https://arxiv.org/abs/https://nyaspubs.onlinelibrary.wiley.com/doi/pdf/10.1111/j.1749-6632.1991.tb32200.x} {https://nyaspubs.onlinelibrary.wiley.com/doi/pdf/10.1111/j.1749-6632.1991.tb32200.x} \BibitemShut {NoStop}%
\bibitem [{\citenamefont {Alford}(2001)}]{Alford2001CSCReview}%
  \BibitemOpen
  \bibfield  {author} {\bibinfo {author} {\bibfnamefont {M.}~\bibnamefont {Alford}},\ }\href {https://doi.org/10.1146/annurev.nucl.51.101701.132449} {\bibfield  {journal} {\bibinfo  {journal} {Annual Review of Nuclear and Particle Science}\ }\textbf {\bibinfo {volume} {51}},\ \bibinfo {pages} {131} (\bibinfo {year} {2001})}\BibitemShut {NoStop}%
\bibitem [{\citenamefont {Alford}\ \emph {et~al.}(2003)\citenamefont {Alford}, \citenamefont {Bowers}, \citenamefont {Cheyne},\ and\ \citenamefont {Cowan}}]{PRDAlfordNonLocking}%
  \BibitemOpen
  \bibfield  {author} {\bibinfo {author} {\bibfnamefont {M.~G.}\ \bibnamefont {Alford}}, \bibinfo {author} {\bibfnamefont {J.~A.}\ \bibnamefont {Bowers}}, \bibinfo {author} {\bibfnamefont {J.~M.}\ \bibnamefont {Cheyne}},\ and\ \bibinfo {author} {\bibfnamefont {G.~A.}\ \bibnamefont {Cowan}},\ }\href {https://doi.org/10.1103/PhysRevD.67.054018} {\bibfield  {journal} {\bibinfo  {journal} {Phys. Rev. D}\ }\textbf {\bibinfo {volume} {67}},\ \bibinfo {pages} {054018} (\bibinfo {year} {2003})},\ \Eprint {https://arxiv.org/abs/hep-ph/0210106} {hep-ph/0210106} \BibitemShut {NoStop}%
\bibitem [{\citenamefont {Schmitt}(2005)}]{Schmitt_spin_one_csc}%
  \BibitemOpen
  \bibfield  {author} {\bibinfo {author} {\bibfnamefont {A.}~\bibnamefont {Schmitt}},\ }\href {https://doi.org/10.1103/PhysRevD.71.054016} {\bibfield  {journal} {\bibinfo  {journal} {Phys. Rev. D}\ }\textbf {\bibinfo {volume} {71}},\ \bibinfo {pages} {054016} (\bibinfo {year} {2005})}\BibitemShut {NoStop}%
\bibitem [{\citenamefont {Sch\"afer}(2006)}]{Schafer:2006ue}%
  \BibitemOpen
  \bibfield  {author} {\bibinfo {author} {\bibfnamefont {T.}~\bibnamefont {Sch\"afer}},\ }\href {https://doi.org/10.1103/PhysRevD.74.054009} {\bibfield  {journal} {\bibinfo  {journal} {Phys. Rev. D}\ }\textbf {\bibinfo {volume} {74}},\ \bibinfo {pages} {054009} (\bibinfo {year} {2006})},\ \Eprint {https://arxiv.org/abs/hep-ph/0606026} {arXiv:hep-ph/0606026} \BibitemShut {NoStop}%
\bibitem [{\citenamefont {Baym}\ \emph {et~al.}(2018)\citenamefont {Baym}, \citenamefont {Hatsuda}, \citenamefont {Kojo}, \citenamefont {Powell}, \citenamefont {Song},\ and\ \citenamefont {Takatsuka}}]{Baym_2018}%
  \BibitemOpen
  \bibfield  {author} {\bibinfo {author} {\bibfnamefont {G.}~\bibnamefont {Baym}}, \bibinfo {author} {\bibfnamefont {T.}~\bibnamefont {Hatsuda}}, \bibinfo {author} {\bibfnamefont {T.}~\bibnamefont {Kojo}}, \bibinfo {author} {\bibfnamefont {P.~D.}\ \bibnamefont {Powell}}, \bibinfo {author} {\bibfnamefont {Y.}~\bibnamefont {Song}},\ and\ \bibinfo {author} {\bibfnamefont {T.}~\bibnamefont {Takatsuka}},\ }\href {https://doi.org/10.1088/1361-6633/aaae14} {\bibfield  {journal} {\bibinfo  {journal} {Reports on Progress in Physics}\ }\textbf {\bibinfo {volume} {81}},\ \bibinfo {pages} {056902} (\bibinfo {year} {2018})}\BibitemShut {NoStop}%
\bibitem [{\citenamefont {Buballa}(2005)}]{Buballa:2005}%
  \BibitemOpen
  \bibfield  {author} {\bibinfo {author} {\bibfnamefont {M.}~\bibnamefont {Buballa}},\ }\href {https://doi.org/https://doi.org/10.1016/j.physrep.2004.11.004} {\bibfield  {journal} {\bibinfo  {journal} {Physics Reports}\ }\textbf {\bibinfo {volume} {407}},\ \bibinfo {pages} {205} (\bibinfo {year} {2005})},\ \Eprint {https://arxiv.org/abs/arXiv:hep-ph/0402234} {arXiv:hep-ph/0402234} \BibitemShut {NoStop}%
\bibitem [{\citenamefont {Song}\ \emph {et~al.}(2019)\citenamefont {Song}, \citenamefont {Baym}, \citenamefont {Hatsuda},\ and\ \citenamefont {Kojo}}]{PRDSongBaymGluonRepulsion}%
  \BibitemOpen
  \bibfield  {author} {\bibinfo {author} {\bibfnamefont {Y.}~\bibnamefont {Song}}, \bibinfo {author} {\bibfnamefont {G.}~\bibnamefont {Baym}}, \bibinfo {author} {\bibfnamefont {T.}~\bibnamefont {Hatsuda}},\ and\ \bibinfo {author} {\bibfnamefont {T.}~\bibnamefont {Kojo}},\ }\href {https://doi.org/10.1103/PhysRevD.100.034018} {\bibfield  {journal} {\bibinfo  {journal} {Phys. Rev. D}\ }\textbf {\bibinfo {volume} {100}},\ \bibinfo {pages} {034018} (\bibinfo {year} {2019})}\BibitemShut {NoStop}%
\bibitem [{\citenamefont {Alford}\ \emph {et~al.}(2008)\citenamefont {Alford}, \citenamefont {Schmitt}, \citenamefont {Rajagopal},\ and\ \citenamefont {Sch\"afer}}]{Alford:2007xm}%
  \BibitemOpen
  \bibfield  {author} {\bibinfo {author} {\bibfnamefont {M.~G.}\ \bibnamefont {Alford}}, \bibinfo {author} {\bibfnamefont {A.}~\bibnamefont {Schmitt}}, \bibinfo {author} {\bibfnamefont {K.}~\bibnamefont {Rajagopal}},\ and\ \bibinfo {author} {\bibfnamefont {T.}~\bibnamefont {Sch\"afer}},\ }\href {https://doi.org/10.1103/RevModPhys.80.1455} {\bibfield  {journal} {\bibinfo  {journal} {Rev. Mod. Phys.}\ }\textbf {\bibinfo {volume} {80}},\ \bibinfo {pages} {1455} (\bibinfo {year} {2008})},\ \Eprint {https://arxiv.org/abs/0709.4635} {arXiv:0709.4635 [hep-ph]} \BibitemShut {NoStop}%
\bibitem [{\citenamefont {Baym}\ \emph {et~al.}(2019)\citenamefont {Baym}, \citenamefont {Furusawa}, \citenamefont {Hatsuda}, \citenamefont {Kojo},\ and\ \citenamefont {Togashi}}]{Baym_2019}%
  \BibitemOpen
  \bibfield  {author} {\bibinfo {author} {\bibfnamefont {G.}~\bibnamefont {Baym}}, \bibinfo {author} {\bibfnamefont {S.}~\bibnamefont {Furusawa}}, \bibinfo {author} {\bibfnamefont {T.}~\bibnamefont {Hatsuda}}, \bibinfo {author} {\bibfnamefont {T.}~\bibnamefont {Kojo}},\ and\ \bibinfo {author} {\bibfnamefont {H.}~\bibnamefont {Togashi}},\ }\href {https://doi.org/10.3847/1538-4357/ab441e} {\bibfield  {journal} {\bibinfo  {journal} {The Astrophysical Journal}\ }\textbf {\bibinfo {volume} {885}},\ \bibinfo {pages} {42} (\bibinfo {year} {2019})}\BibitemShut {NoStop}%
\bibitem [{\citenamefont {Yakovlev}\ and\ \citenamefont {Pethick}(2004)}]{Yakovlev:2004iq}%
  \BibitemOpen
  \bibfield  {author} {\bibinfo {author} {\bibfnamefont {D.~G.}\ \bibnamefont {Yakovlev}}\ and\ \bibinfo {author} {\bibfnamefont {C.~J.}\ \bibnamefont {Pethick}},\ }\href {https://doi.org/10.1146/annurev.astro.42.053102.134013} {\bibfield  {journal} {\bibinfo  {journal} {Ann. Rev. Astron. Astrophys.}\ }\textbf {\bibinfo {volume} {42}},\ \bibinfo {pages} {169} (\bibinfo {year} {2004})},\ \Eprint {https://arxiv.org/abs/astro-ph/0402143} {arXiv:astro-ph/0402143} \BibitemShut {NoStop}%
\bibitem [{\citenamefont {{Page}}\ \emph {et~al.}(2009)\citenamefont {{Page}}, \citenamefont {{Lattimer}}, \citenamefont {{Prakash}},\ and\ \citenamefont {{Steiner}}}]{Page:2009}%
  \BibitemOpen
  \bibfield  {author} {\bibinfo {author} {\bibfnamefont {D.}~\bibnamefont {{Page}}}, \bibinfo {author} {\bibfnamefont {J.~M.}\ \bibnamefont {{Lattimer}}}, \bibinfo {author} {\bibfnamefont {M.}~\bibnamefont {{Prakash}}},\ and\ \bibinfo {author} {\bibfnamefont {A.~W.}\ \bibnamefont {{Steiner}}},\ }\href {https://doi.org/10.1088/0004-637X/707/2/1131} {\bibfield  {journal} {\bibinfo  {journal} {\apj}\ }\textbf {\bibinfo {volume} {707}},\ \bibinfo {pages} {1131} (\bibinfo {year} {2009})},\ \Eprint {https://arxiv.org/abs/0906.1621} {arXiv:0906.1621 [astro-ph.SR]} \BibitemShut {NoStop}%
\bibitem [{\citenamefont {Potekhin}\ \emph {et~al.}(2015)\citenamefont {Potekhin}, \citenamefont {Pons},\ and\ \citenamefont {Page}}]{Potekhin:2015qsa}%
  \BibitemOpen
  \bibfield  {author} {\bibinfo {author} {\bibfnamefont {A.~Y.}\ \bibnamefont {Potekhin}}, \bibinfo {author} {\bibfnamefont {J.~A.}\ \bibnamefont {Pons}},\ and\ \bibinfo {author} {\bibfnamefont {D.}~\bibnamefont {Page}},\ }\href {https://doi.org/10.1007/s11214-015-0180-9} {\bibfield  {journal} {\bibinfo  {journal} {Space Sci. Rev.}\ }\textbf {\bibinfo {volume} {191}},\ \bibinfo {pages} {239} (\bibinfo {year} {2015})},\ \Eprint {https://arxiv.org/abs/1507.06186} {arXiv:1507.06186 [astro-ph.HE]} \BibitemShut {NoStop}%
\bibitem [{\citenamefont {Lattimer}\ \emph {et~al.}(1991)\citenamefont {Lattimer}, \citenamefont {Prakash}, \citenamefont {Pethick},\ and\ \citenamefont {Haensel}}]{Lattimer:1991ib}%
  \BibitemOpen
  \bibfield  {author} {\bibinfo {author} {\bibfnamefont {J.~M.}\ \bibnamefont {Lattimer}}, \bibinfo {author} {\bibfnamefont {M.}~\bibnamefont {Prakash}}, \bibinfo {author} {\bibfnamefont {C.~J.}\ \bibnamefont {Pethick}},\ and\ \bibinfo {author} {\bibfnamefont {P.}~\bibnamefont {Haensel}},\ }\href {https://doi.org/10.1103/PhysRevLett.66.2701} {\bibfield  {journal} {\bibinfo  {journal} {Phys. Rev. Lett.}\ }\textbf {\bibinfo {volume} {66}},\ \bibinfo {pages} {2701} (\bibinfo {year} {1991})}\BibitemShut {NoStop}%
\bibitem [{\citenamefont {Brown}\ \emph {et~al.}(2018)\citenamefont {Brown}, \citenamefont {Cumming}, \citenamefont {Fattoyev}, \citenamefont {Horowitz}, \citenamefont {Page},\ and\ \citenamefont {Reddy}}]{Brown:2017gxd}%
  \BibitemOpen
  \bibfield  {author} {\bibinfo {author} {\bibfnamefont {E.~F.}\ \bibnamefont {Brown}}, \bibinfo {author} {\bibfnamefont {A.}~\bibnamefont {Cumming}}, \bibinfo {author} {\bibfnamefont {F.~J.}\ \bibnamefont {Fattoyev}}, \bibinfo {author} {\bibfnamefont {C.~J.}\ \bibnamefont {Horowitz}}, \bibinfo {author} {\bibfnamefont {D.}~\bibnamefont {Page}},\ and\ \bibinfo {author} {\bibfnamefont {S.}~\bibnamefont {Reddy}},\ }\href {https://doi.org/10.1103/PhysRevLett.120.182701} {\bibfield  {journal} {\bibinfo  {journal} {Phys. Rev. Lett.}\ }\textbf {\bibinfo {volume} {120}},\ \bibinfo {pages} {182701} (\bibinfo {year} {2018})},\ \Eprint {https://arxiv.org/abs/1801.00041} {arXiv:1801.00041 [astro-ph.HE]} \BibitemShut {NoStop}%
\bibitem [{\citenamefont {Cumming}\ \emph {et~al.}(2017)\citenamefont {Cumming}, \citenamefont {Brown}, \citenamefont {Fattoyev}, \citenamefont {Horowitz}, \citenamefont {Page},\ and\ \citenamefont {Reddy}}]{Cumming:2016weq}%
  \BibitemOpen
  \bibfield  {author} {\bibinfo {author} {\bibfnamefont {A.}~\bibnamefont {Cumming}}, \bibinfo {author} {\bibfnamefont {E.~F.}\ \bibnamefont {Brown}}, \bibinfo {author} {\bibfnamefont {F.~J.}\ \bibnamefont {Fattoyev}}, \bibinfo {author} {\bibfnamefont {C.~J.}\ \bibnamefont {Horowitz}}, \bibinfo {author} {\bibfnamefont {D.}~\bibnamefont {Page}},\ and\ \bibinfo {author} {\bibfnamefont {S.}~\bibnamefont {Reddy}},\ }\href {https://doi.org/10.1103/PhysRevC.95.025806} {\bibfield  {journal} {\bibinfo  {journal} {Phys. Rev. C}\ }\textbf {\bibinfo {volume} {95}},\ \bibinfo {pages} {025806} (\bibinfo {year} {2017})},\ \Eprint {https://arxiv.org/abs/1608.07532} {arXiv:1608.07532 [astro-ph.HE]} \BibitemShut {NoStop}%
\bibitem [{\citenamefont {Holt}\ \emph {et~al.}(2020)\citenamefont {Holt}, \citenamefont {Kawaguchi},\ and\ \citenamefont {Kaiser}}]{Holt:2019bah}%
  \BibitemOpen
  \bibfield  {author} {\bibinfo {author} {\bibfnamefont {J.~W.}\ \bibnamefont {Holt}}, \bibinfo {author} {\bibfnamefont {M.}~\bibnamefont {Kawaguchi}},\ and\ \bibinfo {author} {\bibfnamefont {N.}~\bibnamefont {Kaiser}},\ }\href {https://doi.org/10.3389/fphy.2020.00100} {\bibfield  {journal} {\bibinfo  {journal} {Front. in Phys.}\ }\textbf {\bibinfo {volume} {8}},\ \bibinfo {pages} {100} (\bibinfo {year} {2020})},\ \Eprint {https://arxiv.org/abs/1912.06055} {arXiv:1912.06055 [nucl-th]} \BibitemShut {NoStop}%
\end{thebibliography}%
\end{document}